\documentclass[12pt,a4paper]{article}
\usepackage[left=2.3cm, right=2.3cm, top=2.54cm]{geometry}

\usepackage{amsmath}
\usepackage{amsfonts}
\usepackage{graphicx}
\usepackage[font=small,labelfont=bf]{caption}
\usepackage{epstopdf}
\usepackage{authblk}
\usepackage{subcaption}
\usepackage{multirow}
\usepackage{tikz}
\usepackage{hyperref}
\usetikzlibrary{positioning}
\usetikzlibrary{backgrounds}

\def\GLnZ{$GL(n,\mathbb{Z})$ }

\def\h11{h^{1,1}}
\def\cpp{C\texttt{++} }
\def\D491{\Delta^{\circ}_{491}}

\newcommand{\gappeq}{\mathrel{\rlap {\raise.5ex\hbox{$>$}}
{\lower.5ex\hbox{$\sim$}}}}
\newcommand{\lappeq}{\mathrel{\rlap{\raise.5ex\hbox{$<$}}
{\lower.5ex\hbox{$\sim$}}}}

\title{\textbf{Estimating Calabi-Yau Hypersurface and Triangulation 
Counts
with Equation Learners}}
\author[]{Ross Altman}
\author[]{Jonathan Carifio}
\author[]{James Halverson}
\author[]{Brent D. Nelson}

\date{}

\affil[]{Department of Physics, Northeastern University\\Boston, MA 02115, USA}

\begin{document}
\maketitle

\vspace{1cm}

\begin{abstract}
We provide the first estimate of the number of fine, regular, star triangulations of the four-dimensional reflexive polytopes, as classified by Kreuzer and Skarke (KS). This provides an upper
bound on the number of Calabi-Yau threefold hypersurfaces
in toric varieties. The estimate is performed with
deep learning, specifically the novel equation learner (EQL)
architecture. We demonstrate that EQL networks accurately
predict numbers of triangulations far beyond the $\h11$
training region, allowing for reliable extrapolation. 
We estimate that number of triangulations in the KS dataset is $10^{10,505}$, dominated by the polytope with the highest $h^{1,1}$ value.
 \end{abstract}

\clearpage

\tableofcontents

%\section{Introduction}

\clearpage 
\section{Motivation}
\label{sec:intro}

Triangulations and Calabi-Yau manifolds are objects
of intrinsic
mathematical interest in combinatorics and algebraic 
geometry, respectively. In the former case, elementary
operations on triangulations such as flips may be used
to determine when two triangulations are canonically
related to one another, or to populate large ensembles
of triangulations. Similarly, elementary topology
changing operations on Calabi-Yau manifolds such
as flop or conifold transitions may relate
two such manifolds to one another by finite distance
motion in metric moduli space; these may be
utilized to generate large ensembles of geometries. 
In the case of Calabi-Yau hypersurfaces in toric varieties,
Calabi-Yau data is encoded in the structure of a triangulation,
and operations on the triangulation often induce topological
transitions in the Calabi-Yau hypersurface. The transitions 
between these objects naturally generate a large network
that encodes ensemble structure.

%\vspace{.3cm}
In string theory, Calabi-Yau manifolds are some of the simplest
and best-studied backgrounds on which to compactify the
extra dimensions of space while preserving supersymmetry in
four dimensions \cite{Candelas:1985en}. There, flop \cite{Aspinwall:1993nu} and 
conifold \cite{Strominger:1995cz} transitions 
induce
space-time topology change in a physically consistent manner
and may be utilized to generate large ensembles of string
compactifications; generalized fluxes \cite{Dasgupta:1999ss,Giddings:2001yu} also give rise to large ensembles \cite{Bousso:2000xa}.
The associated collection of four-dimensional
effective potentials and
metastable vacua form the so-called string landscape,
which is central to understanding the implications of string theory
for particle physics and cosmology.

A natural direction is to understand cosmological
dynamics on the landscape and associated mechanisms
for vacuum selection. Such dynamics,
together with statistical properties of vacua \cite{Douglas:2003um},
could then lead to concrete statistical predictions
for physics in four dimensions. Global structures in the landscape, such as large networks,
may play a role in the dynamics. For example,
dynamical vacuum selection \cite{Carifio:2017nyb}
 on a network of string geometries \cite{Halverson:2017ffz}
 in a well-studied bubble cosmology \cite{Garriga:2005av}
selects models with large numbers of gauge groups and axions,
as well as strong coupling.
However, concrete studies of the landscape are difficult
due to its enormity \cite{Ashok:2003gk,Denef:2004ze,Taylor:2015xtz,Taylor:2015ppa,Halverson:2017ffz,Taylor:2017yqr}, computational complexity \cite{Denef:2006ad,Cvetic:2010ky,Bao:2017thx,Denef:2017cxt,Halverson:2018cio}, and undecidability \cite{Cvetic:2010ky}. It is therefore natural to expect that,
in addition to the formal progress that is clearly required,
data science techniques such as supervised machine learning
will be necessary to understand the landscape; see 
for initial works \cite{1706.02714,Ruehle2017,Carifio:2017nyb,Krefl2017} in this directions and \cite{Carifio2018,1809.02547,Bull2018,Wang2018,Hashimoto2018,Liu2017,
1805.12153} for
additional promising results.

%\vspace{.3cm}
One concrete goal is to understand the full ensemble of Calabi-Yau threefolds, network-like structures induced by
transitions between them, and associated implications for
cosmological dynamics and vacuum selection in compatifications of string theory.
However, this is far out of reach currently for
reasons of enormity and complexity.

Instead, in this paper we take a modest but necessary first step in this direction. We will estimate the number of Calabi-Yau
threefold hypersurfaces in toric varieties, which are one
of the most-studied ensembles in string theory. Each such
manifold is naturally associated to a 4d reflexive polytope,
which were classified in a seminal work \cite{Kreuzer:2000xy}
by Kreuzer and Skarke. 
Determining a Calabi-Yau from its corresponding polytope requires the
specification of a fine, regular, star triangulation (FRST) of
the polytope, and our main result is that
\begin{equation}
n_{\text{FRST}} \simeq 10^{10,505.2\pm292.6},
\end{equation}
where $n_{\text{FRST}}$ is the number of FRSTs arising in
the Kreuzer-Skarke ensemble. We obtain this estimate with
deep learning, specifically a novel neural network architecture
known as an equation learner (EQL) \cite{1610.02995}, which
we demonstrate is significantly better at extrapolating
triangulation predictions to large $\h11$ than a standard
feed-forward neural network. In particular, by demonstrating
accurate extrapolation to high $\h11$ in the analogous
problem for 3d reflexive polytopes, we lend
credibility to the 4d prediction, which requires extrapolation
significantly beyond the regime in which training data is available. Predicting
$n_\text{FRST}$
provides an estimated upper bound on the number of Calabi-Yau threefold
hypersurfaces in toric varieties. 

\vspace{.3cm}
Given these accurate predictions of $n_\text{FRST}$, it is interesting
to study the interpretability of the predictions made by the neural network.
Sometimes refered to as intelligible artificial intelligence, interpretability
is a major current goal of machine learning research, and it is one
of the reasons for developing the EQL architecture; see \cite{Carifio:2017bov} for the related idea of conjecture generation, by which interpretable
numerical decisions may be turned into rigorous results. In the EQL context,
the idea is to mimic what happens in natural sciences such as physics,
where a physical phenomenon is often described in terms of an interpretable
function that allows for understanding and generalization. Accordingly,
by utilizing simpler functions than standard architectures, EQLs 
in principle increase the likelihood of intepretability. 

In the case of $n_\text{FRST}$, we found that an EQL that utilizes
quadratic functions makes accurate predictions; see Sections \ref{sec:ML}
and \ref{sec:3d} for quadratic functions associated to trained EQLs.
By studying an associated heat map of coefficients, we demonstrate that
some variables and cross-correlations are clearly of more importance than
others, but the existence of a large number of warm spots suggest that
many variables matter, which makes interpretability difficult. This could
be an artifact of having used a quadratic function, which may be suboptimal,
but it could also be the case that there is no simple interpretation
of $n_\text{FRST}$ predictions. That is, sufficiently complex phenomena
in complex systems may not admit descriptions in terms of simple equations
in human-understandable variables.

\vspace{.3cm}
This paper is organized as follows: In Section 2, we give an overview of our approach for estimating the number of FRSTs in the 4d Kreuzer-Skarke database. In Section 3 we discuss the method by which we classified the 3d facets of the 4d reflexive polytopes, the results of this classification, and our success in obtaining FRTs of these facets. In Section 4 we discuss our unsuccessful initial machine learning attempts, our input features, and our successful EQL model. In Section 5, we perform a similar analysis for the 3d reflexive polytopes and show that a model of a similar architecture is able to extrapolate far outside its training range.

\section{Approach}

Batyrev has shown~\cite{Batyrev:1994hm} that a hypersurface in a toric variety can be chosen to be Calabi-Yau if the object underlying the construction of the variety, a lattice polytope, obeys the condition of reflexivity.

A reflexive polytope $\Delta$ is defined as the convex hull of a set of points $\{v\} \in \mathbb{Z}^{n}$ whose dual polytope
\begin{equation}
\Delta^{\circ} = \{w \in \mathbb{Z}^{n} \, | \, w \cdot v \ge -1 \;\;\; \forall v \in \Delta\}
\label{dualpoly} \end{equation}
is itself a lattice polytope. In four dimensions, there are $473,800,776$ reflexive polytopes, as classified by Kreuzer and Skarke~\cite{Kreuzer:2000xy}. The ambient 4d space described by these polytopes is generally a singular toric variety. A proper Calabi-Yau manifold will therefore be a hypersurface in a suitably de-singularized ambient space. The process of de-singularization is equivalent to obtaining a fine, regular, star triangulation (FRST) of the 4d reflexive polytope. 

Crucially, there are typically very many inequivalent FRSTs for a given polytope. While software packages such as {\tt PALP}~\cite{Kreuzer:2002uu} and {\tt TOPCOM}~\cite{Rambau:TOPCOM-ICMS:2002} can, in principle, produce all possible triangulations of a given polytope; in practice, such a request becomes increasingly problematic computationally as the overall size of the polytope grows. A good proxy for this computational load is the value of the topological quantity $h^{1,1}$ associated with the polytope. For example, in~\cite{Altman:2014bfa}, all triangulations of all 4d reflexive polytopes with $h^{1,1} \leq 6$ were obtained. The nearly 652,000 unique triangulations required approximately 120,000 core-hours of processing time to obtain. The computational burden can be mitigated somewhat by exploiting the reflexivity property of these polytopes~\cite{Long:2014fba}. Even with this assistance, however, obtaining more than a single, canonical triangulation for a given polytope (in a reasonable computational time) becomes difficult for $h^{1,1} \gappeq 10$~\cite{Demirtas:2018akl}. It would appear, therefore, that an enumeration of the unique triangulations in the KS 4d reflexive polytope dataset must remain unobtainable, barring dramatic advances in computational power or a vastly superior triangulation algorithm. It is this impasse which forms the motivation for the current work.

There is reason to believe that application of machine learning techniques may allow for an estimate of the total number of triangulations in the KS dataset to be achieved. A proof of principle already exists in the KS set of 3d reflexive polytopes, of which there are 4319. In~\cite{Halverson:2016tve}, the number of FRSTs for many of the 3d reflexive polytopes was computed, and an estimation of the total number was obtained, using standard triangulation techniques. Soon thereafter, another estimate was obtained using supervised machine learning. Specifically, a decision-tree regression model was trained on known results, and used to estimate the number of FRSTs for the remaining cases~\cite{Carifio:2017bov}. The results were in good agreement with the original estimate of~\cite{Halverson:2016tve}. The objective of the current work was to obtain such an estimate for the 4d reflexive polytopes via similar techniques.

As was done for the 3d case, the method employed here is to estimate the number of FRSTs of a given polytope as the product of the number of fine, regular triangulations (FRTs) of each of its facets. However, counting the number of vertices in the 4d reflexive dataset (which equals the number of facets by duality) shows that there are over 7.5 billion facets. In order to better get a handle on this set, as well as avoid unnecessary computational repetition, the first step was a classification of these 3d facets. The procedure for doing so is the subject of Section~\ref{sec:facet}.

With this classification in hand, we then explicitly computed the number of FRTs for as many of the facets as possible. Due to computational constraints, this consisted of only 1.03\% of the total -- primarily consisting of facets that first appear in dual polytopes at low values of $\h11$. The data associated with these explicitly-triangulated facets then became the training data for supervised machine learning. Ultimately, a neural network was employed to construct a model which predicts the natural logarithm of the number of FRTs of each facet. From this we arrive at an estimate of the total number of FRSTs of the 4d reflexive polytopes. The machine learning techniques, and FRST estimate, are the subject of Section~\ref{sec:ML}.

Any such machine learning estimate will obviously have limitations, and ours come from the two main issues that we faced. First, as noted above, we were only able to obtain known results for 1.03\% of cases, meaning that it was necessary to estimate the value for nearly 99\%. Second, as the facets that were able to be triangulated were necessarily ones with fewer triangulations, this problem involved extrapolation to output values beyond the range seen in the training set. However, as we will discuss below, we believe that our estimate is sound. The equation learning (EQL) neural network architecture that we ultimately employed was developed especially for extrapolation outside of a given training set. By withholding the subset of our known results that first appeared at $\h11 \geq 12$ when training, we were able to see that our model extrapolated with stable results over a large number of cases as the number of triangulations increased. We comment further on these limitations, and the potential overcounting associated with our approach, in Section~\ref{sec:conclusion}.

\section{Facet Classification}

In this section we review methods that easily determine when two facets
are equivalent, and then apply those methods to the classification of
facets in 4d reflexive polytopes.

\label{sec:facet}

\subsection{Distinguishing unique facets}

We must distinguish between three-dimensional (3d) facets of the four-dimensional (4d) reflexive polytopes, since individual facets may appear in multiple polytopes. In particular, we must determine under what conditions two facets should be considered to be equivalent, and then find a method to determine whether the conditions are satisfied.

For their classification of the 3d and 4d reflexive polyhedra~\cite{Kreuzer:2000xy}, Kreuzer and Skarke defined a normal form, with the property that two reflexive polyhedra have the same normal form if and only if they are related by a \GLnZ transformation. While powerful, this normal form has the restriction that it can only be used on full-dimensional polytopes.

To circumvent this issue, and thereby be able to use the normal form, we employ the method of Grinis and~Kasprzyk, as described in Section 3.2 of~\cite{1301.6641}. 
The origin is interior to every reflexive polytope, and thus we know that none of the facets contain the origin. Thus, for each facet $F$, we constructed the associated subcone defined by $C_{F} = \textrm{conv}(F\cup\lbrace0\rbrace)$. The subcone is a full-dimensional polytope, and so its normal form can be computed. Additionally, as the origin is the sole interior lattice point of a reflexive polytope, the origin is the only lattice point in $C_{F} \setminus F$. This also means that we need not worry about lattice translations when comparing subcones. As the origin is fixed under any \GLnZ transformation, two subcones $C_{F_{1}}$ and $C_{F_{2}}$ have the same normal form if and only if their associated facets $F_{1}$ and $F_{2}$ are related by a \GLnZ transformation.

%\subsubsection{Example}

As an example, consider the following two facets $F_{1}$ and $F_{2}$, both of which appear as dual facets to the same $h^{1,1}=2$ polytope (given by POLYID 21 in the ToricCY database~\cite{Altman:2014bfa}). Each of the facets are the convex hulls of four vertices, as below:
\begin{eqnarray}
F_{1} &=& \textrm{conv}(\{\{-1,0,0,0\},\{-1,0,0,1\},\{-1,0,1,0\},\{-1,1,0,0\}\}) \\
F_{2} &=& \textrm{conv}(\{\{-1,0,0,1\},\{-1,0,1,0\},\{1,0,0,0\},\{2,-1,-1,-1\}\})\, .
\label{facetex}
\end{eqnarray}
Adding the origin to each facet, we obtain the associated subcones
\begin{eqnarray}
C_{F_{1}} &=& \textrm{conv}(\{\{0,0,0,0\},\{-1,0,0,0\},\{-1,0,0,1\},\{-1,0,1,0\},\{-1,1,0,0\}\}) \\
C_{F_{2}} &=& \textrm{conv}(\{\{0,0,0,0\}\{-1,0,0,1\},\{-1,0,1,0\},\{1,0,0,0\},\{2,-1,-1,-1\}\})
\label{subconesex}
\end{eqnarray}
Computing the normal form for each subcone, we find that
\begin{equation}
\textrm{NF}(C_{F_{1}}) = \textrm{NF}(C_{F_{2}}) = \{\{0,0,0,0\},\{1,0,0,0\},\{0,1,0,0\},\{0,0,1,0\},\{0,0,0,1\}\})\, .
\end{equation}
We see that the two subcones have the same normal form, and thus the facets are equivalent. Removing the origin, we see that both facets are equivalent to the 3d polytope with vertices $\{\{1,0,0,0\},\{0,1,0,0\},\{0,0,1,0\},\{0,0,0,1\}\}$.

As discussed in Section~\ref{sec:class} below, this particular facet, known as the standard 3-simplex, is the most common facet among the 4d reflexive polytopes. It accounts for $20.45\%$ of all facets, while no other facet accounts for more than $8.57\%$.

\subsection{Performing the classification}

The primary practical challenge of the facet classification was the volume of the 4d polytope dataset itself. In order to have a consistent direction for our computation, we worked through the polytopes in order of increasing $\h11(X)$, where $X$ is an associated Calabi-Yau hypersurface. As it is the dual polytope that is triangulated when constructing this Calabi-Yau, we identified the facets of the dual polytope in each case. Hence, for each $\h11$ value, we identified the facets that appeared in each dual polytope, keeping both a master list of unique facets as well as recording which facets appeared in each dual polytope (and with which multiplicity).

Computation was done on Northeastern University's Discovery cluster using the SLURM workload manager. As identifying the dual facets is an independent computation for each polytope, we were able to use distributed computing to decrease our real-world running time by several orders of magnitude. Filtering and removal of duplicate facets was performed after identification had been completed for each $\h11$ value.
The dual and normal form computations were done using a \cpp lattice polytope implementation of our own creation which used the {\tt PALP} source code for many of its underlying calculations. This offered significantly faster performance than the LatticePolytope class in {\tt Sage}, allowing the classification to be finished on the timespan of a few weeks.

\subsection{Classification results}
\label{sec:class}

Using the above method for the facets of every 4d reflexive polytope, we found that there are a total of $45,990,557$ unique 3d facets, which is an order of magnitude less than the number of polytopes themselves. This is approximately $0.6\%$ of the total number of facets, which is $7,471,985,487$.
We found that a relatively small number of facets accounted for the majority of the facets that appear across the polytopes, with the most common facet accounting for just over $20\%$ of the total. 

\begin{center}
\begin{figure}[th]
\begin{subfigure}{.5\textwidth}
	\includegraphics[scale=0.45]{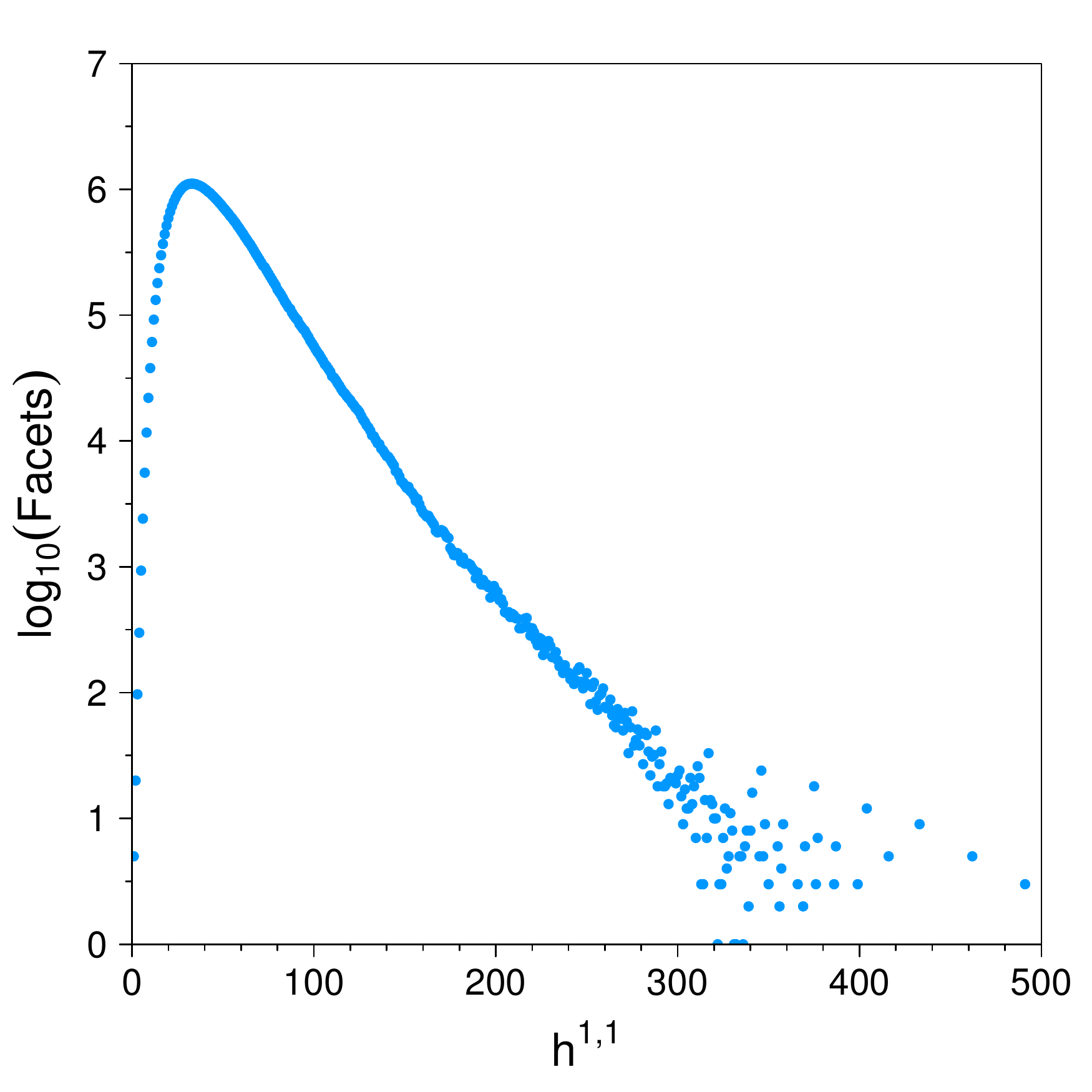}
\end{subfigure}
\begin{subfigure}{.5\textwidth}
	\includegraphics[scale=0.45]{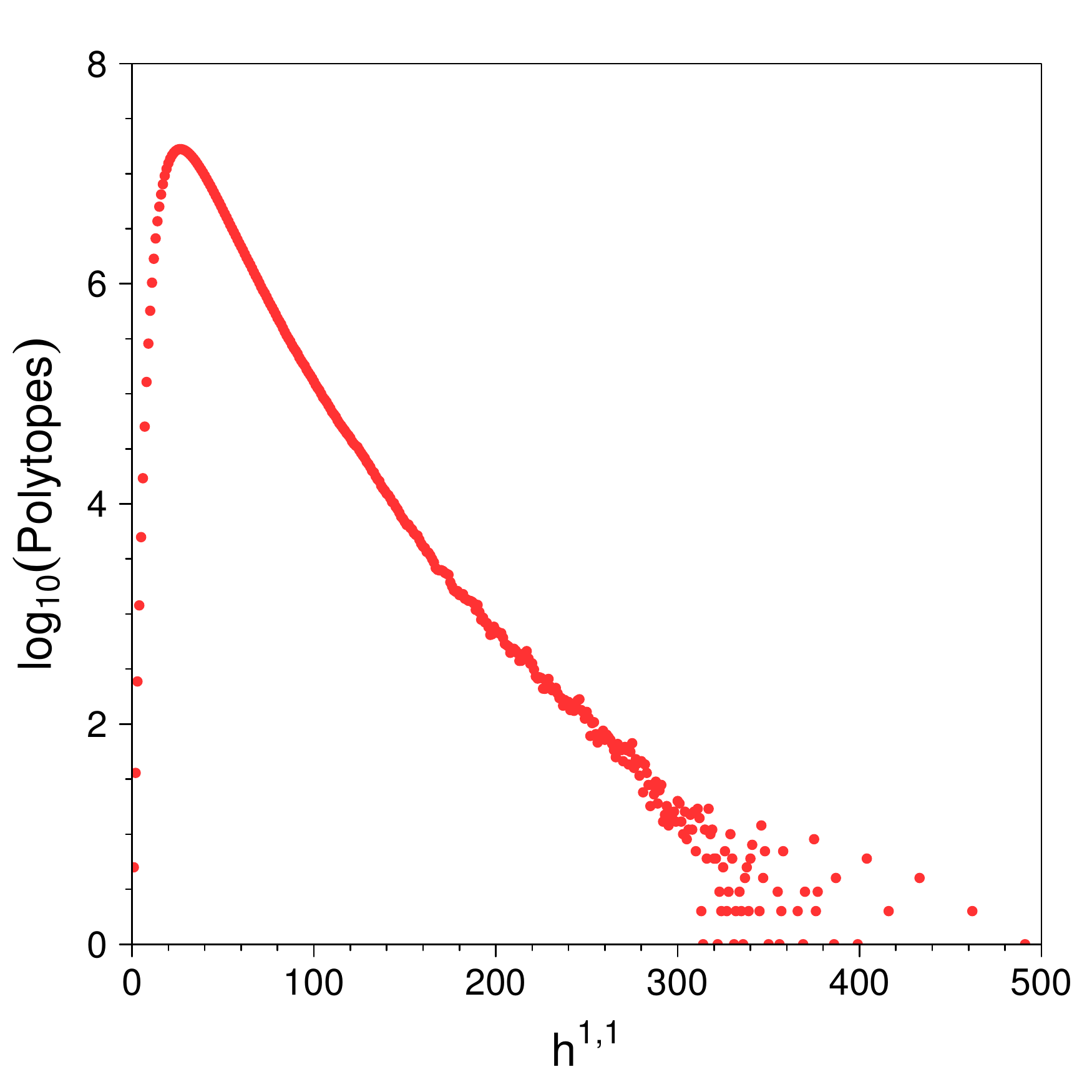}
\end{subfigure}
\caption{\textbf{(Left)} The logarithm of the number of new facets at each $\h11$ value. \textbf{(Right)} The logarithm of the number of reflexive polytopes at each $\h11$ value.}
\label{fig:Distributions}
\end{figure}
\end{center}

As we performed the classification procedure, we recorded the number of  facets which appeared for the first time at each $\h11$ value; i.e.,
the same facet may appear in many different polytopes with different values
of $\h11$, and we recorded the smallest such value.  The distribution of these new  facets per $\h11$ value is shown in the left panel of Figure~\ref{fig:Distributions}, which should be compared to the distribution of the number of polytopes of a given $\h11$ value, which is given in the right panel of Figure~\ref{fig:Distributions}. The shape of the two distributions are very similar, albeit decreased by around an order of magnitude. We note that the two distributions have different peaks: the number of reflexive polytopes peaks at $\h11=27$, while the number of new facets peaks at $\h11=33$.

\begin{center}
\begin{figure}[htbp!]
\begin{subfigure}{.5\textwidth}
	\includegraphics[scale=0.45]{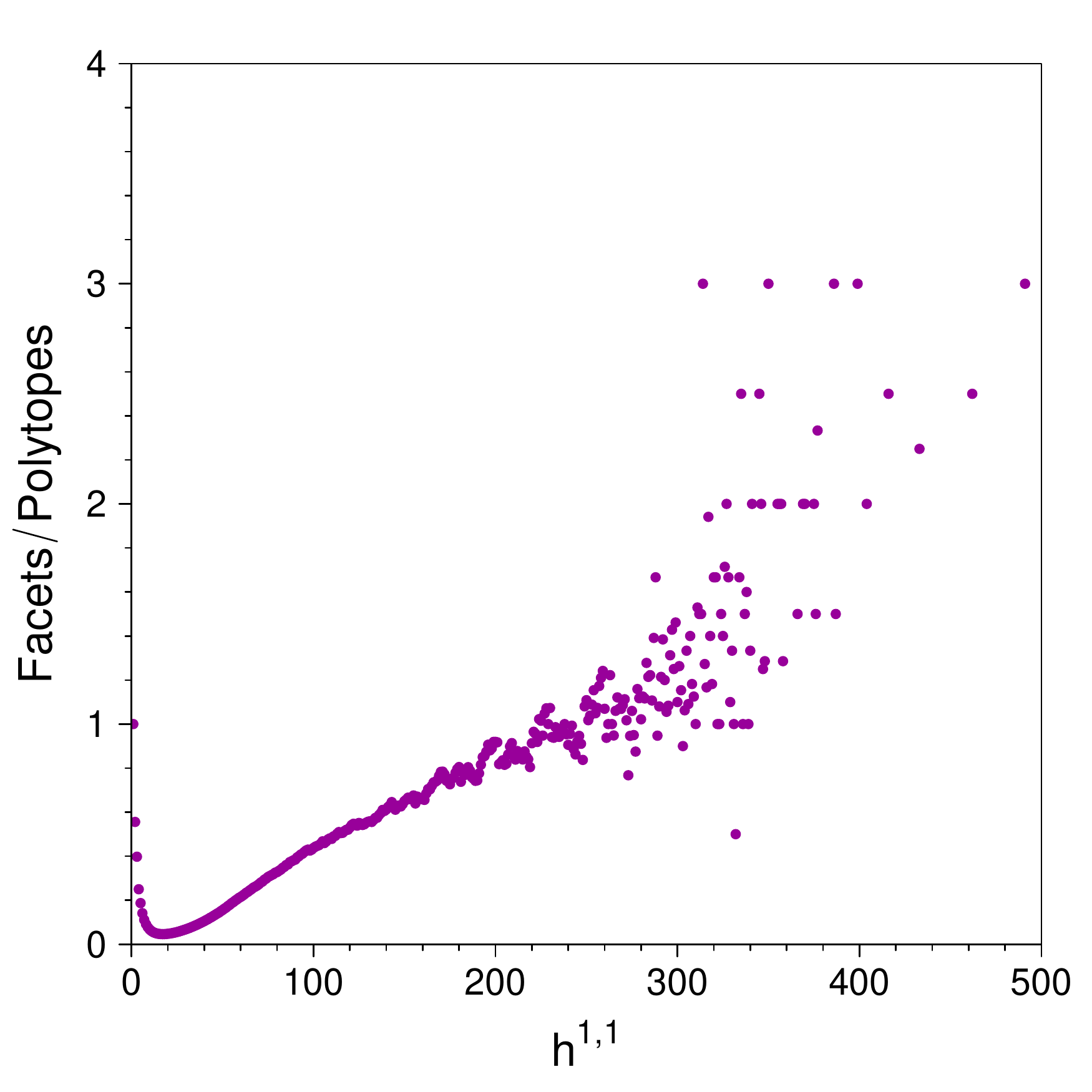}
\end{subfigure}
\begin{subfigure}{.5\textwidth}
	\includegraphics[scale=0.45]{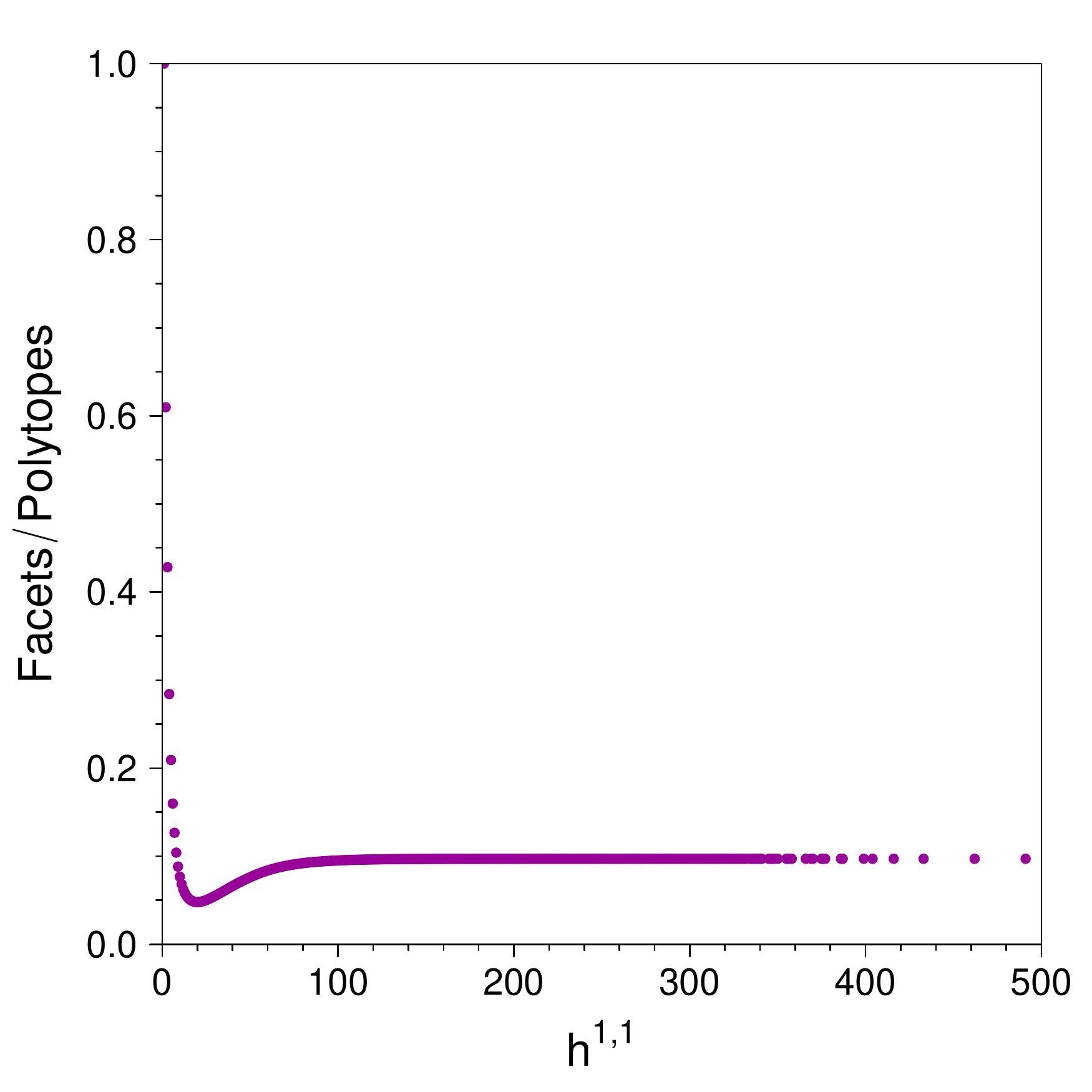}
\end{subfigure}
\caption{\textbf{(Left)} The number of new facets at each $\h11$ value, as a fraction of the number of polytopes at that $\h11$. \textbf{(Right)} The total number of  facets found through each $\h11$ value, as a fraction of the total number of polytopes up to that point.}
\label{fig:RatioDistributions}
\end{figure}
\end{center}

Along with the cumulative distribution, we also show these distributions as fractions of the relevant numbers of polytopes in Figure~\ref{fig:RatioDistributions}. In the left panel, we show the number of new facets that appear at a given $\h11$ value, as a fraction of the total number of polytopes at that $\h11$ value. The scatter that emerges for $\h11 \gappeq 250$ represents the relatively small number of polytopes at these rather large $\h11$ values. Coupled with the information contained in Figure~\ref{fig:Distributions}, it is clear that the overwhelming majority of facets are encountered well before $\h11 \simeq 250$, though (as we will see), the number of triangulations are dominated by the outliers in the very large $\h11$ bins. 

Finally, the right panel of Figure~\ref{fig:RatioDistributions} shows the number of new facets at a given $\h11$ value, normalized by the \textbf{cumulative} number of polytopes to that particular $\h11$ value. We see that after the peak in the right panel of Figure~\ref{fig:Distributions}, the ratio quickly saturates to a value of approximately~0.1. This is nothing more than the ratio of total unique facets found ($47\times 10^6$) to the number of 4d reflexive polytopes in the KS database ($470 \times 10^6$).
%%%

\begin{center}
\begin{figure}[th!]
\begin{subfigure}{.5\textwidth}
	\includegraphics[scale=0.45]{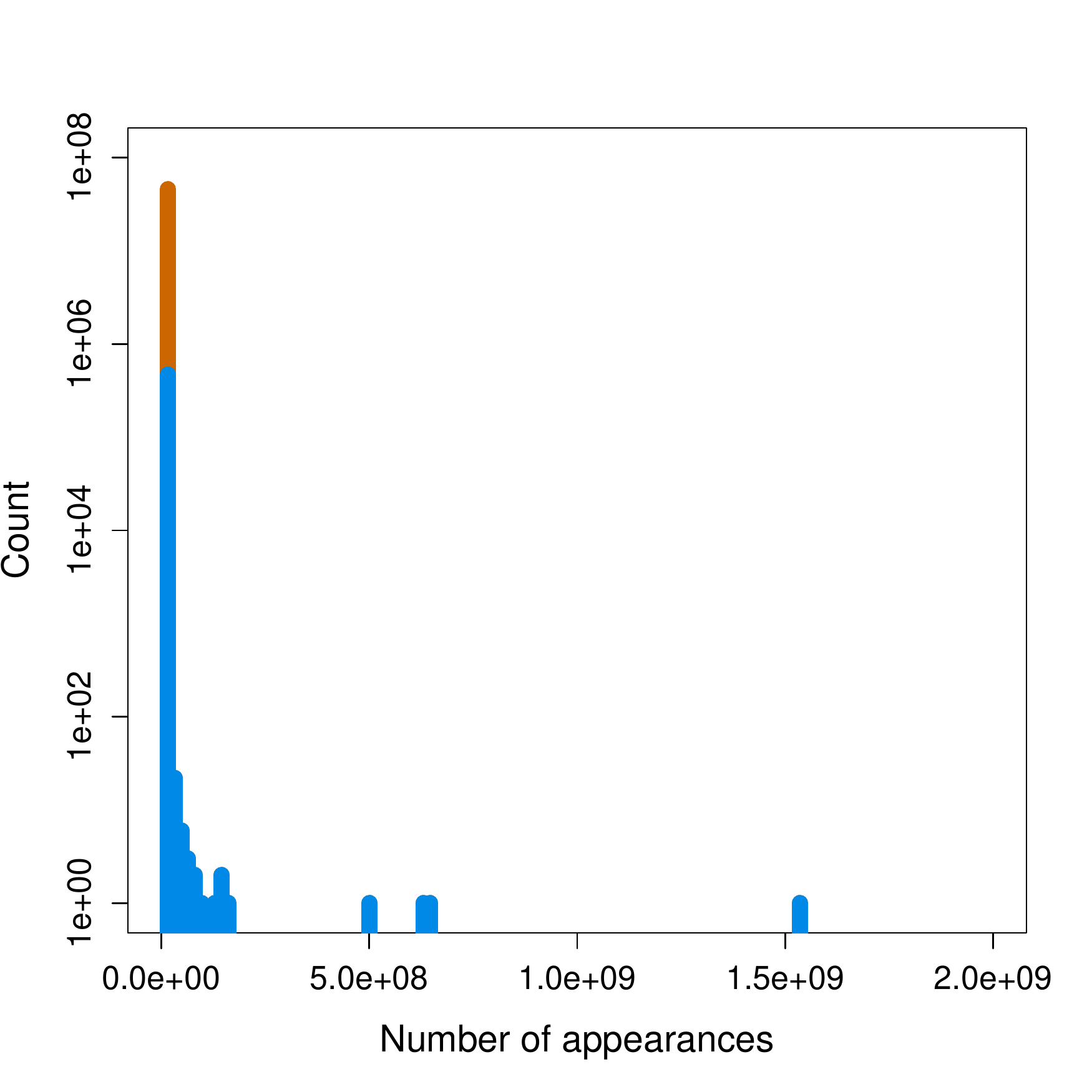}
\end{subfigure}
\begin{subfigure}{.5\textwidth}
	\includegraphics[scale=0.45]{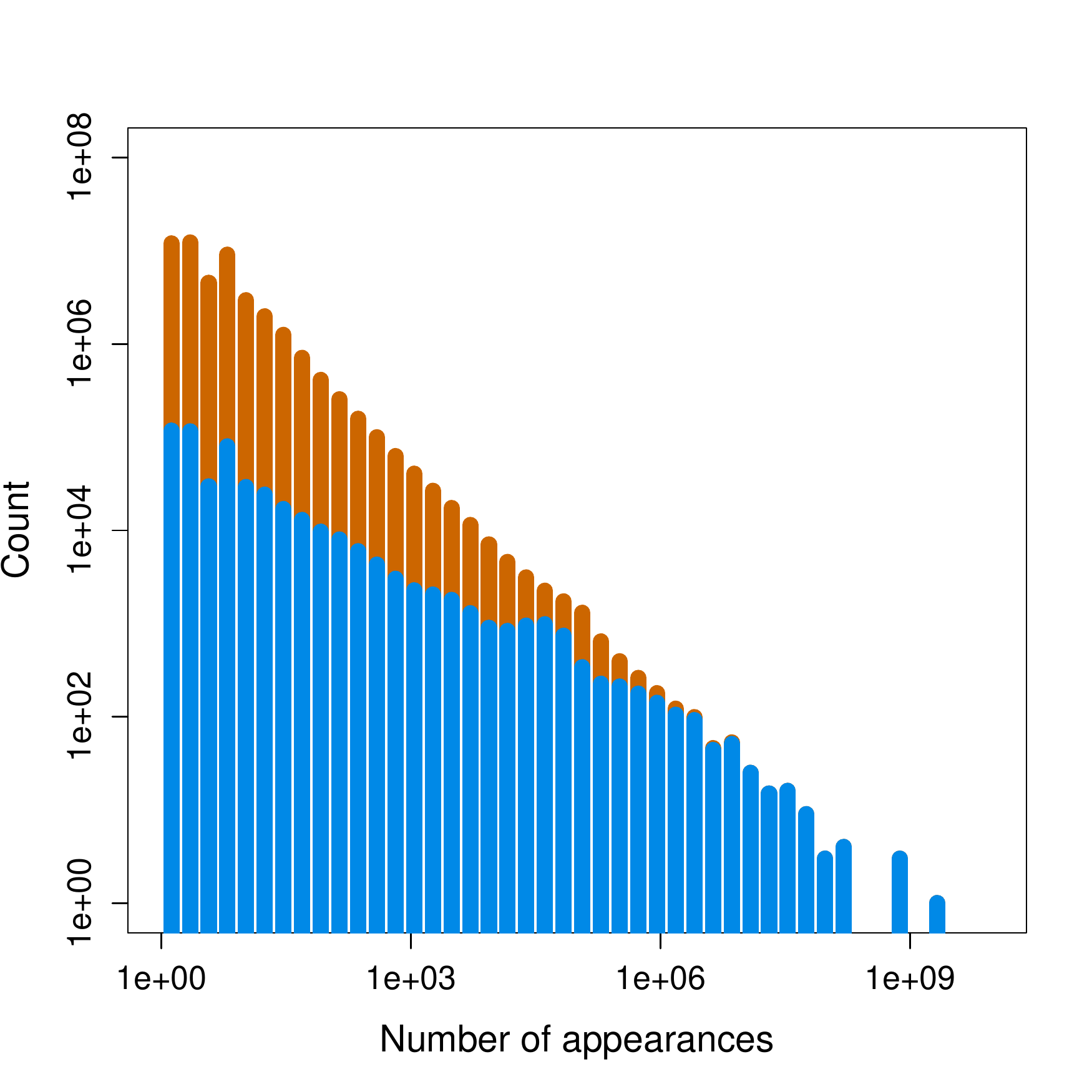}
\end{subfigure}
\caption{Histograms of the frequencies of facet appearances, with bins spaced linearly \textbf{(left panel)}, and logarithmically \textbf{(right panel)}. The orange bars represent the total number of facets, while the blue bars show the amount for which the number of FRTs is explicitly computed, and thus known.}
\label{fig:FrequencyHistogram}
\end{figure}
\end{center}

In understanding the reliability of our estimation methods, it is important to ask how often the unique facets that we have enumerated actually appear in the 4d polytopes in the KS database. This is the subject of Figure~\ref{fig:FrequencyHistogram}, in which we show a histogram of the number of appearances of a given facet.

One can see from the left panel of Figure~\ref{fig:FrequencyHistogram} that a small number of facets dominate the database, with the frequency counts decreasing linearly on a logarithmic scale (the right panel of Figure~\ref{fig:FrequencyHistogram}). The 100~most common facets account for 74\% of the total number. As mentioned in Section~\ref{sec:intro}, we were able to achieve all possible FRTs for a very small set of the total number of 3d facets, but these cases represent over 88\% of facets by appearance. This is indicated by the blue shading in both panels of Figure~\ref{fig:FrequencyHistogram}.

%\subsubsection{The standard 3-simplex}
\begin{center}
\begin{figure}[th!]
\begin{subfigure}{.5\textwidth}
	\includegraphics[scale=0.45]{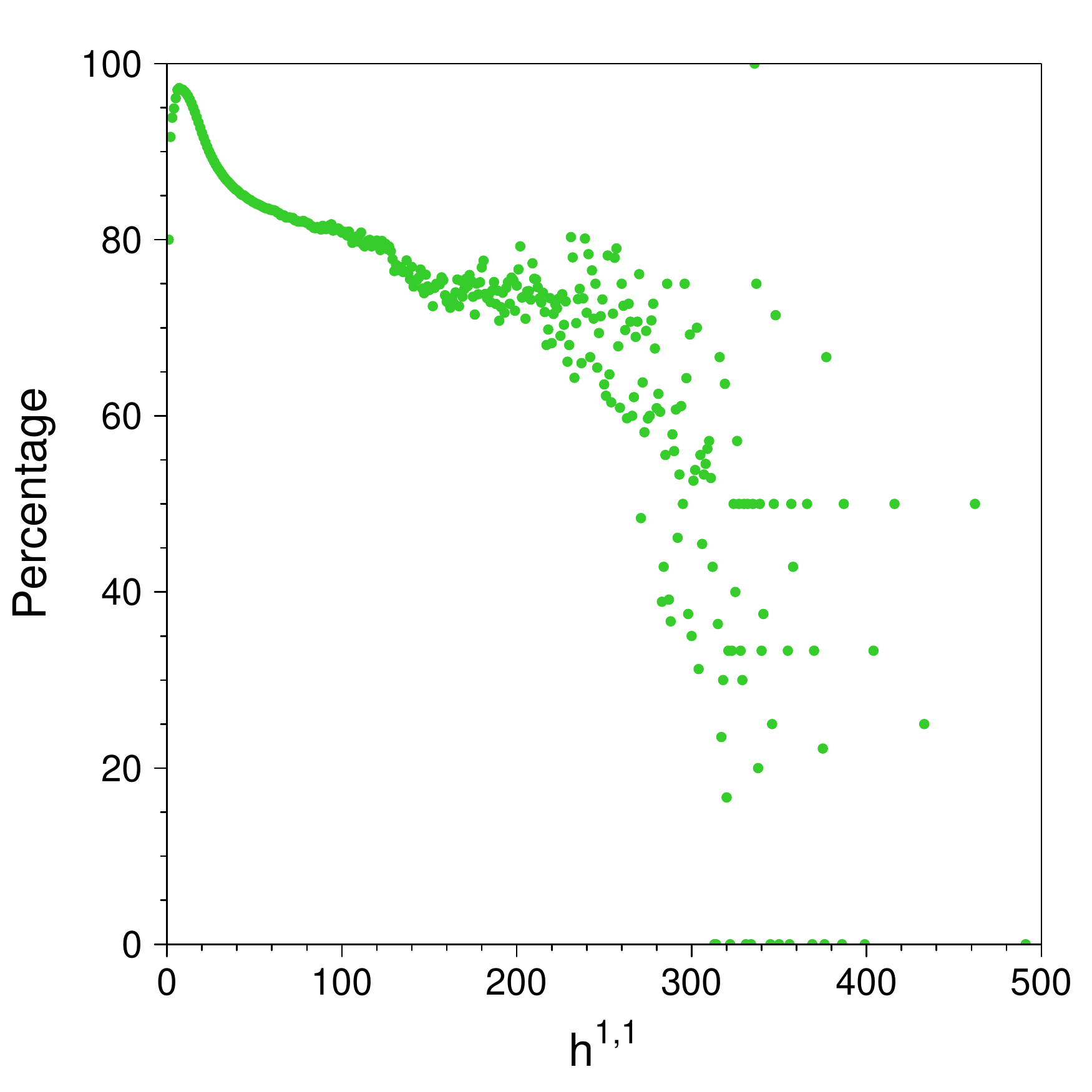}
\end{subfigure}
\begin{subfigure}{.5\textwidth}
	\includegraphics[scale=0.45]{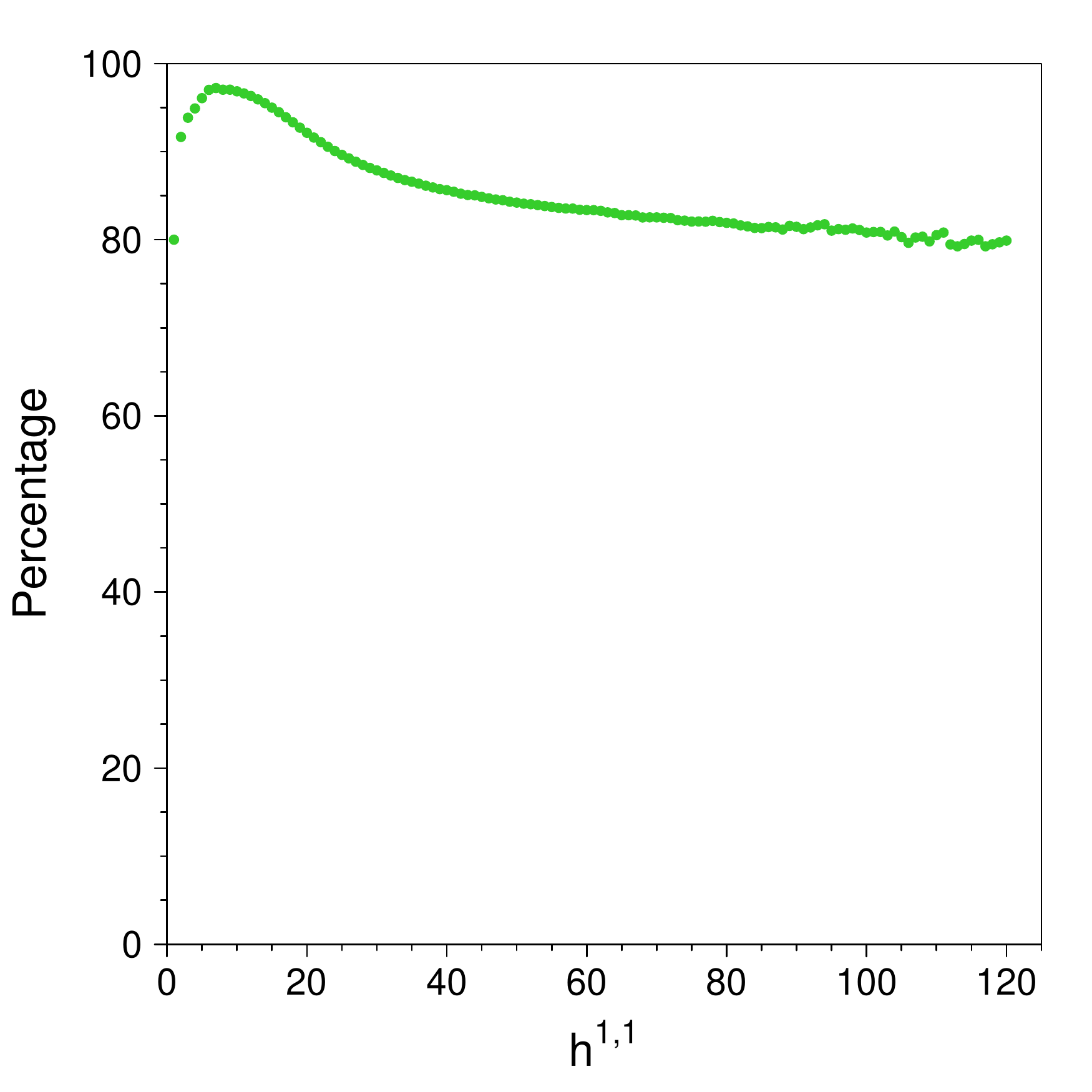}
\end{subfigure}
\caption{\textbf{(Left)} The percentage of dual polytopes that contain S3S at each $\h11$ value. \textbf{(Right)} The same graph, truncated at $\h11\leq120$. We note that the erratic portion of the Figure~\ref{fig:S3S} for $\h11 \gappeq 120$ accounts for fewer than 1 million polytopes. The truncated graph (Figure \ref{fig:S3S}b) shows more clearly the prevalence of S3S.}
\label{fig:S3S}
\end{figure}
\end{center}

The most common facet is the 3d polyhedron known as the standard 3-simplex (S3S), which has normal form vertices
\begin{equation}
\{\{1,0,0,0\},\{0,1,0,0\},\{0,0,1,0\},\{0,0,0,1\}\}\, .
\label{S3S}
\end{equation}
The S3S appears a total of $1,528,150,671$ times in the database, accounting for $20.45\%$ of all facets, and appears in 87.8\% of the polytopes. It first appears as a dual facet at $\h11=1$ and maintains a consistent presence through the database, as evidenced by the graphs in Figure~\ref{fig:S3S}. However, the S3S, being itself a simplex with no interior points, has only one FRT, and thus has essentially no effect on the total number of polytope FRSTs.

\section{Machine Learning Numbers of Triangulations}
\label{sec:ML}

In supervised machine learning (often called simply \textit{supervised learning}), the output of the machine learning algorithm is a function, commonly called a \textit{model}, which takes a specified set of inputs and produces a unique output value. This function contains numerical parameters, or \textit{weights}, which are adjusted by the machine learning algorithm. The algorithm is trained on a set of input $\rightarrow$ output pairs, and attempts to minimize a predetermined loss function by adjusting the weights of the function.

In this section we will estimate the number of FRSTs of 4d reflexive 
polytopes by using supervised learning to predict the number of FRTs of the
facets.
For this application, the input is a set of features which describe a facet, and the desired output is the number of FRTs for that facet. We describe in this section the features which we used as input and the structure of the model that we obtained via supervised learning.

\subsection{Training data}

The first step in our machine learning process was to generate data on which to train a model. In this case, this meant it was necessary to triangulate as many of the 3d facets as possible, in order to obtain the best possible training set.

\begin{table}[t]
\begin{center}
\begin{tabular} {|c|c|c|c|}
\hline
\textbf{$\h11$} & \textbf{Facets} & \textbf{Triangulated} & \textbf{\% Triangulated} \\ \hline
$1-11$ & 142,257 & 142,257 & 100\% \\ \hline
12 & 92,178 & 92,162 & 99.983\% \\ \hline
13 & 132,153 & 108,494 & 82.097\% \\ \hline
14 & 180,034 & 124,700 & 69.625\% \\ \hline
15 & 236,476 & 3,907 & 1.652\% \\ \hline
$>15$ & 45,207,459 & 1,360 & 0.003\% \\ \hline
Total & 45,990,557 & 472,896 & 1.028\% \\ \hline
\end{tabular}
\end{center}
\caption{Dual facet FRT numbers obtained, binned by the first $\h11$ value at which they appear.}
\label{table:facet}
\end{table}

Due to computational restrictions, we were only able to find the actual number of FRTs for $472,880$ of the facets, which represents $1.03\%$ of the total. These consist almost entirely of facets which first appear in the dual polytope at relatively low $\h11$ values. Table~\ref{table:facet} shows our progress. As one can see, $\h11=14$ was the last value at which we were able to obtain an appreciable fraction of the real FRT values. While the triangulated set may constitute only a small fraction of the unique facets, they account for over 88\% of all facet appearances.

To obtain the FRT values, we first computed the 2-skeleton of each facet using our \cpp code. The n-skeleton of a lattice polytope consists of all points which are not interior to any (n+1)-dimensional face. We then passed the 2-skeleton to the {\tt TOPCOM} executable {\tt points2nfinetriangs} in order to obtain the FRT number. As with the classification process, we used Northeastern's Discovery cluster to run simultaneous calculations for several hundred facets at a time.

When training with this data, and throughout our process of finding a good model, we organized the facets by the first $\h11$ at which they appeared as a dual facet, which we will refer to from here on as the $\h11$ value of the facet. Our reason for this was that there is a rough correlation between this $\h11$ value and the number of triangulations. This is reflected in Table~\ref{table:facet}: as $\h11$ increased, fewer cases were able to complete as the facets became more complex. For this reason, the training set for each model only contained facets with $\h11$ up to some maximum value (typically 11). This allowed us to see how well the model would extrapolate to FRT numbers higher than it had seen, something that would prove to be an issue throughout our attempts.

\subsection{Initial attempts}

Our first attempt at obtaining a model was to follow the same pattern used for the 3d polytopes in Section~3 of~\cite{Carifio:2017bov}. As was done in the 3d case, for each polytope we constructed the 4-tuple $(n_{p}, n_{i}, n_{b}, n_{v})$, consisting of the number of points, interior points, boundary points, and vertices, respectively. Our metric for determining the model's success was the mean absolute percent error (MAPE), which is defined as
\begin{equation} 
\textrm{MAPE} = \frac{100}{n} \times \sum_{i=1}^{n}\left|\frac{A_{i}-P_{i}}{A_{i}}\right|\, ,
\label{MAPE}
\end{equation}
where $n$ is the number of data points, and $P_{i}$ and $A_{i}$ are the predicted and actual values for the output, which here is $\textrm{ln}(N_{FRT})$ for the $i^{th}$ facet.

We trained models using the same four algorithms as discussed in~\cite{Carifio:2017bov}: Linear Discriminant Analysis (LDA), k-Nearest Neighbors (KNNR), Classification and Regression Trees (CART), and Naive Bayes (NB). Of these, the CART algorithm gave impressive performance on the training set, as well as on the test data for the $\h11$ values that it had trained on. However, its ability to extrapolate to higher $\h11$ values was poor. As the majority of the facets lie at higher $\h11$ values, the inability to extrapolate well was unacceptable.

As an example, we show here the results from one of these attempts. The model is an ExtraTreesRegressor model from the {\tt scikit-learn} Python package, using 35~estimators. We trained the model on a data set consisting of 60\% of the known values from the range $5 \le \h11 \le 10$. On the training set, the model achieved a MAPE of 5.723. It performed similarly on the test set consisting of the other 40\% of these values, with a MAPE of 5.821. If we could be confident of a similar accuracy for the rest of the facets, this could be an acceptable model. However, when we used the model to predict on facets from higher $\h11$ values whose FRT numbers are known, we obtained the results shown in Table~\ref{table:ETRPredictions}.

\begin{table}[t]
\begin{center}
\begin{tabular}{|c|c|c|c|}
\hline
\textbf{$\h11$} & \textbf{MAPE} & \textbf{Actual mean} & \textbf{Predicted mean}  \\ \hline
11 & 6.566 & 9.582 & 9.189 \\ \hline
12 & 9.065 & 10.882 & 9.903 \\ \hline
13 & 11.566 & 11.755 & 10.067 \\ \hline
14 & 17.403 & 12.638 & 10.179 \\ \hline
\end{tabular}
\end{center}
\caption{Prediction results for $\textrm{ln}(N_{FRT})$, using the ExtraTreesRegressor model, for $\h11$ values outside of its training region.}
\label{table:ETRPredictions}
\end{table}

As one can see, the MAPE value consistently increases with the $\h11$ value, indicating that the model is less accurate farther away from its training region. Further, the rightmost two columns illustrate the problem with this model: it is under-predicting on facets with a higher number of triangulations. The model never predicts a $\textrm{ln}(N_{FRT})$ value greater than 12.467, which is similar to 12.595, the highest value in its training set.

This example is emblematic of the problems we faced with traditional machine learning. Models either fit the training data poorly, or had trouble extrapolating outside of their training regions. The primary difference between the 3d case (where this simple approach worked), and the 4d case, is the inability to generate a representative training set. In the 3d case, the number of FRTs for all facets that appear up through $\h11=22$ were triangulated, in a set where the maximum value of $\h11$ is 35. Conversely, in the 4d case, the number of FRTs was only obtained for all facets up through $\h11=11$, and for a majority of cases up to $\h11=14$, in a data set where the maximum $\h11$ value is 491.

\subsection{Neural networks}

Given the poor performance of traditional supervised machine learning as described above, we shifted our focus to artificial neural networks. In particular, a feed-forward network seemed the most suited to our purposes.

We recall briefly here the definitions of a neuron and a neural network. A \textit{neuron} is a function $f(\sum_{i}^{k}w_{i} \cdot x_{i} + b)$ whose argument $x_{i}$ is the input. The value of $f$ is called the \textit{output}. The parameters $w_{i}$ are called \textit{weights}, $b$ is called the \textit{bias} (or \textit{offset}), and $\cdot$ represents the appropriate tensor contraction. The function $f$ itself is called the \textit{activation function} and the choice of $f$ is one of the characteristics that define the neuron.

A \textit{neural network} is a (finite) directed graph, each node of which is a neuron. For each arrow in the graph, the output of the neuron at the tail is used as the input for the neuron at the head. Conceptually, a neural network is organized into \textit{layers}, such that there are no connections between nodes in the same layer. The set of nodes whose arguments  explicitly involve the original input data is known as the \textit{input layer}, while the set whose output involves the actual output data is called the \textit{output layer}. All other layers are referred to as \textit{hidden layers}.

In the general case, a neural network may contain cycles, allowing for complicated connections between layers. In a \textit{feed-forward} neural network, there are no cycles, and so information moves only in one direction. Cycles in a neural network allow for the current output to be influenced by the previous output. As the number of triangulations of a given facet is independent of the number for other facets, we can restrict ourselves to the feed-forward case.

We initially tried applying a feed-forward network to our 4-tuples of data, but despite trying a variety of architectures faced similar issues as in the previous subsection, with models consistently underpredicting outside of the training regime. To aid our networks in their prediction attempts, additional input data was generated. In addition to the numbers of points, interior points, boundary points, and vertices, the following quantities were computed for each facet to be used as inputs to the neural network:
\begin{itemize}
\item{The number of points in the 1- and 2-skeletons}
\item{The first $\h11$ value at which the facet appears in a dual polytope}
\item{The number of faces}
\item{The number of edges}
\item{The number of flips of a seed triangulation of the 2-skeleton. Two triangulations differ by a flip if one can be obtained from the other by removing one edge and inserting another.}
\item{Several quantities obtained from a single fine, regular triangulation of the facet:
	\begin{itemize}
	\item{The total numbers of 1-, 2-, and 3-simplices in the triangulation}
	\item{The numbers of 1- and 2-simplices in the triangulation, without accounting for redundancy between higher-dimensional simplices}
	\item{The numbers of 1- and 2-simplices shared between $N$ 2- and 3-simplices, respectively, for $N$ up to 5}
	\end{itemize}
	}
\end{itemize}

As as example we consider the facet $F$ with normal form is given by
\begin{eqnarray} 
 \{\{0,0,0,0\},\{1,0,0,0\},\{1,2,0,0\},\{0,0,1,0\},\{0,1,1,0\}, \nonumber \\ \quad \{0,0,0,1\},\{3,3,-1,-1\},\{3,2,-2,0\},\{0,1,0,1\}\}\, . \label{exF}
\end{eqnarray}
This facet has 11 lattice points, of which 10 lie on the boundary (leaving 1 in the interior). Of these 10 boundary points, 8 are vertices. The 1- and 2-skeletons of this facet contain 9 and 10 points, respectively. It first appears in a dual polytope with $\h11=8$, and has 10 faces and 16 edges. Its flip graph has 13 nodes. Obtaining one FRT for this facet from {\tt TOPCOM}, we find that it contains 72 1-simplices, 48 2-simplices, and 12 3-simplices. Of the 1-simplices, there were 29 unique ones, while there were 32 unique 2-simplices. There were 13 1-simplices shared between two 2-simplices, 6 shared between three, 3 shared between 4, and 1 shared between 5. There were 16 2-simplices shared between two 3-simplices, with none shared between more than two. Hence the input vector describing this facet was
\begin{equation}
(1,10,8,9,10,8,10,16,13,72,48,12,29,32,13,6,3,1,16,0,0,0)\, .
\label{exinput}
\end{equation}

Adding this additional data to our original inputs improved results, but not to the point of satisfaction. As an example of the struggles faced by a traditional neural network, we show here the results from one such model. This neural network has two hidden layers, each with 30 nodes. The first layer has a sigmoid activation function, while the second has a $\tanh$ activation function. The final layer is a rectified linear unit (ReLU), meaning that its activation function $f_{\rm act}$ is equal to the positive part of its argument:
\begin{equation}
f_{\rm act}(x) = \textrm{max}(0,x)\, .
\label{RELU}
\end{equation}
A ReLU unit was chosen for the final layer as we want $\textrm{ln}(N_{FRT})$ to be positive.

\begin{figure}
\begin{center}
\begin{subfigure}{.3\textwidth}
	\includegraphics[scale=0.27]{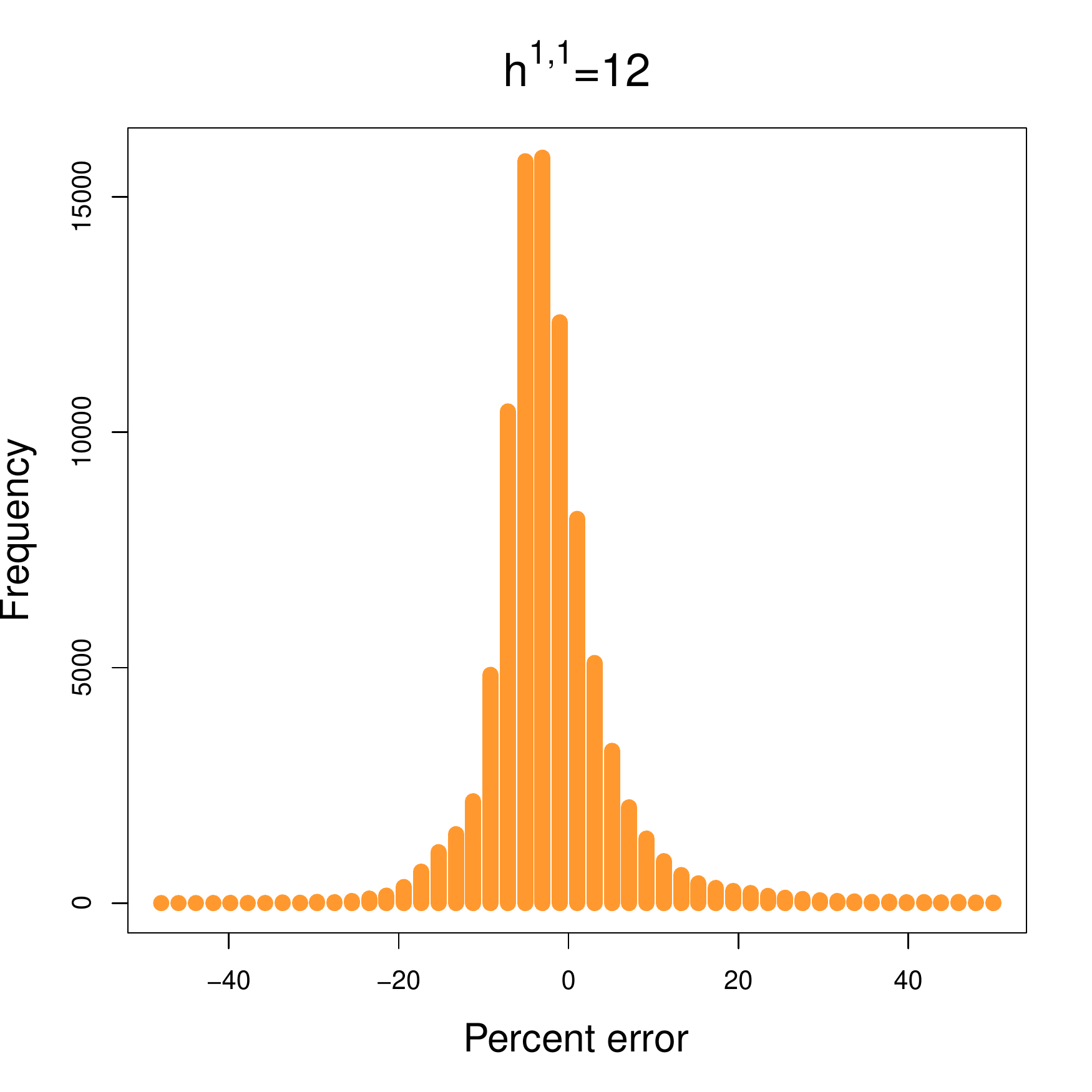}
\end{subfigure}
\begin{subfigure}{.3\textwidth}
	\includegraphics[scale=0.27]{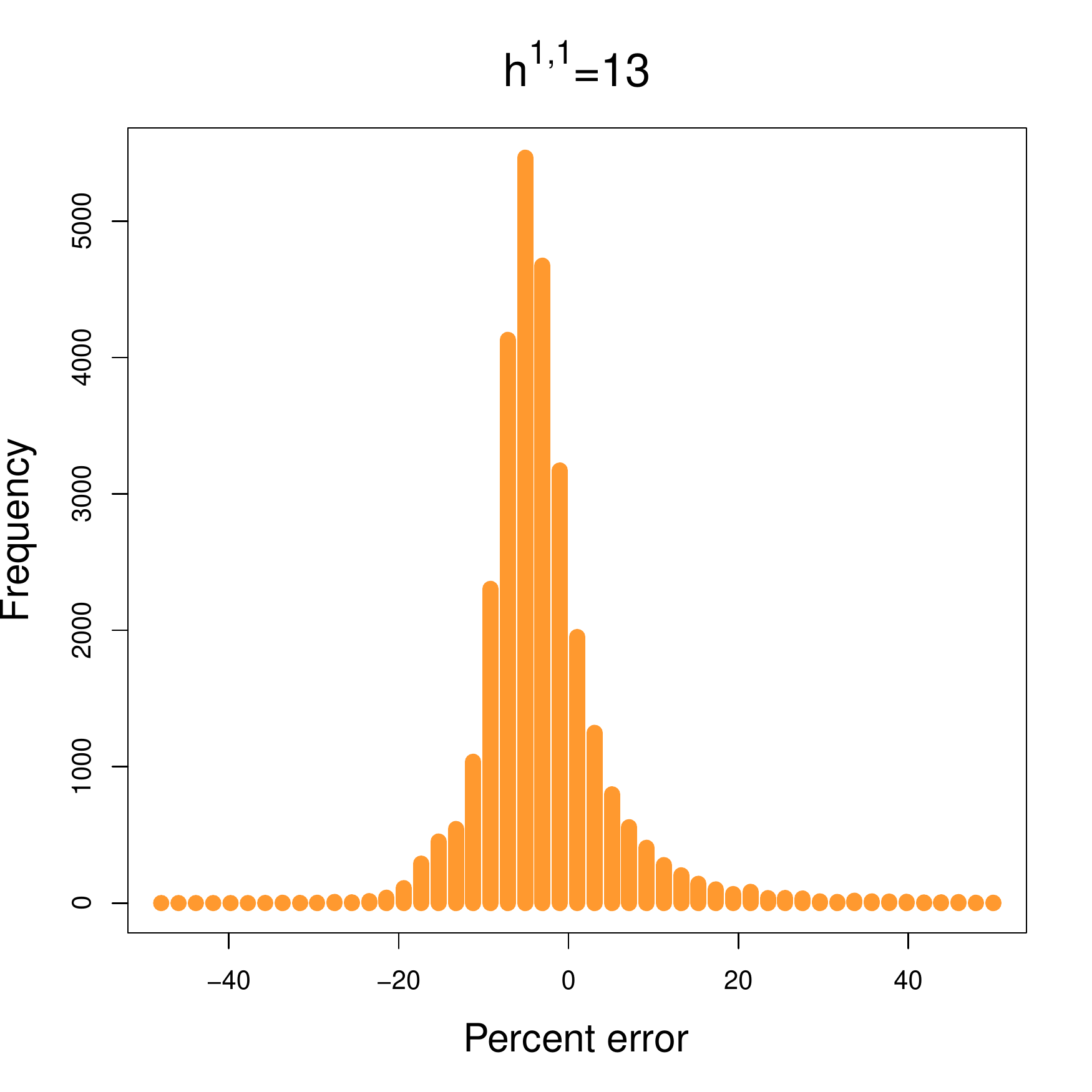}
\end{subfigure}
\begin{subfigure}{.3\textwidth}
	\includegraphics[scale=0.27]{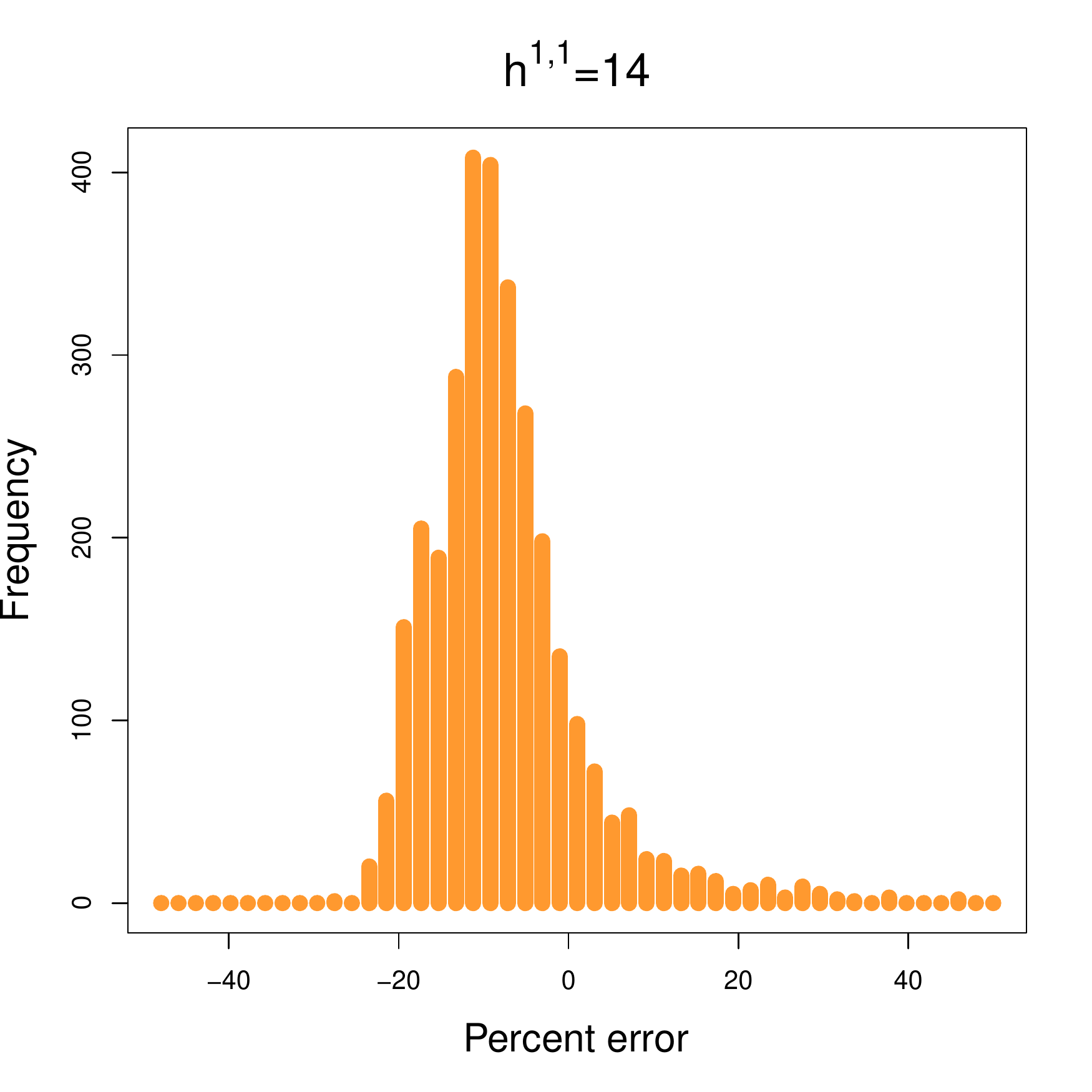}
\end{subfigure}
\end{center}
\caption{Histograms of the percent error of the feed-forward neural network's predictions in the extrapolation region. 
Compare to extrapolations in the EQL network in Figure \ref{fig:PEhistogramsEQL}.}
\label{fig:PEhistograms}
\end{figure}

The model was trained on a data set consisting of an equal number of randomly chosen points from $6 \le \h11 \le 11$ to avoid biasing the model towards higher $\h11$ values. The model performed well on the test set of values from the same $\h11$, with a MAPE of 6.304. However, when we evaluated the model's progress on higher $\h11$ values, our results were as shown in Table \ref{table:NNExample}. This model performs much better than the ExtraTreesRegressor outside of the training region, with a MAPE of 10.915 at $\h11=14$. However, the MAPE values are increasing with $\h11$, and the mean predicted values are falling behind the true mean values, leading one to believe that the model is underpredicting at higher $\h11$. This belief is confirmed by examining the histograms of the percent error values in the extrapolation range, shown in Figure \ref{fig:PEhistograms}. As $\h11$ increases, the percent error distribution skews to the left, indicating that the model is not keeping up with the increasing values. Hence this model is not suitable for extrapolation to the higher $\h11$ values which make up the bulk of our data set.

\begin{table}[t]
\begin{center}
\begin{tabular}{|c|c|c|c|}
\hline
\textbf{$\h11$} & \textbf{MAPE} & \textbf{Mean value} & \textbf{Predicted mean}  \\ \hline
12 & 5.904 & 10.882 & 10.324 \\ \hline
13 & 6.550 & 11.755 & 10.753 \\ \hline
14 & 10.915 & 12.638 & 11.094 \\ \hline
\end{tabular}
\end{center}
\caption{Prediction results for $\textrm{ln}(N_{FRT})$, using the traditional neural network, for $\h11$ values outside of its training region.}
\label{table:NNExample}
\end{table}

\subsection{The EQL architecture}

In this section we will instead utilize an equation learner (EQL) architecture, which was first introduced in \cite{1610.02995}, and
will find that it has significantly better extrapolation ability. 

Each layer in an EQL network consists of two stages: a standard linear stage followed by a non-linear stage. The non-linear stage essentially replaces the activation function in a standard neural network layer, but differs significantly in that it changes the shape of the linear stage's output tensor.
Like any feed-forward neural network layer, an EQL layer accepts some number of input values $n_{i}$, and outputs some number $n_{o}$ of output values. The linear stage maps the $n_{i}$-dimensional input $x$ to an intermediate $n_{m}$-dimensional representation $z$ via an affine transformation. That is,
\begin{equation}
 z = Wx + b
 \end{equation}
for some weight matrix $W \in \mathbb{R}^{n_{m} \times n_{i}}$ and some bias vector $b \in \mathbb{R}^{n_{m}}$.

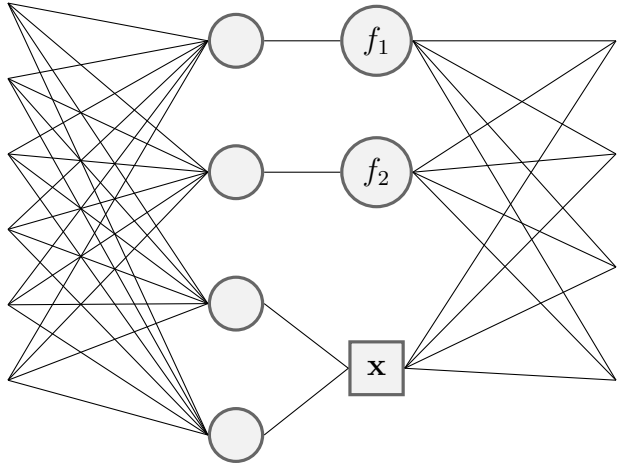
\begin{figure}
\begin{center}
\begin{tikzpicture}[
roundnode/.style={circle, draw=black!60, fill=black!5, very thick, minimum size=7mm},
squarednode/.style={rectangle, draw=black!60, fill=black!5, very thick, minimum size=7mm},
]

%Nodes
\node[roundnode]	(L1)					{};
\node[roundnode]	(R1)	[right=of L1]	{$f_{1}$};
\node[roundnode]	(L2)	[below=of L1]	{};
\node[roundnode]	(R2)	[right=of L2]	{$f_{2}$};
\node[roundnode]	(L3)	[below=of L2]	{};
\node[roundnode]	(L4)	[below=of L3]	{};
\node[squarednode]	(X)		[below= 1.75cm of R2]	{\textbf{x}};
\coordinate			(I1) at (-3,0.5);
\coordinate			(I2) at (-3,-0.5);
\coordinate			(I3) at (-3,-1.5);
\coordinate			(I4) at (-3,-2.5);
\coordinate			(I5) at (-3,-3.5);
\coordinate			(I6) at (-3,-4.5);
\coordinate			(O1) at (5,0);
\coordinate			(O2) at (5,-1.5);
\coordinate			(O3) at (5,-3);
\coordinate			(O4) at (5,-4.5);

%Lines
\draw[-] (L1.east) -- (R1.west);
\draw[-] (L2.east) -- (R2.west);
\draw[-] (L3.east) -- (X.west);
\draw[-] (L4.east) -- (X.west);

% To unseen input
\draw[-] (I1) -- (L1.west);
\draw[-] (I1) -- (L2.west);
\draw[-] (I1) -- (L3.west);
\draw[-] (I1) -- (L4.west);
\draw[-] (I2) -- (L1.west);
\draw[-] (I2) -- (L2.west);
\draw[-] (I2) -- (L3.west);
\draw[-] (I2) -- (L4.west);
\draw[-] (I3) -- (L1.west);
\draw[-] (I3) -- (L2.west);
\draw[-] (I3) -- (L3.west);
\draw[-] (I3) -- (L4.west);
\draw[-] (I4) -- (L1.west);
\draw[-] (I4) -- (L2.west);
\draw[-] (I4) -- (L3.west);
\draw[-] (I4) -- (L4.west);
\draw[-] (I5) -- (L1.west);
\draw[-] (I5) -- (L2.west);
\draw[-] (I5) -- (L3.west);
\draw[-] (I5) -- (L4.west);
\draw[-] (I6) -- (L1.west);
\draw[-] (I6) -- (L2.west);
\draw[-] (I6) -- (L3.west);
\draw[-] (I6) -- (L4.west);

% To unseen output
\draw[-] (O1) -- (R1.east);
\draw[-] (O1) -- (R2.east);
\draw[-] (O1) -- (X.east);
\draw[-] (O2) -- (R1.east);
\draw[-] (O2) -- (R2.east);
\draw[-] (O2) -- (X.east);
\draw[-] (O3) -- (R1.east);
\draw[-] (O3) -- (R2.east);
\draw[-] (O3) -- (X.east);
\draw[-] (O4) -- (R1.east);
\draw[-] (O4) -- (R2.east);
\draw[-] (O4) -- (X.east);

%\draw[fill=gray!50!white (-2,0) rectangle (3,5);

\end{tikzpicture}
\caption{A representation of a simple EQL layer with $n_{m} = 4$ and $n_{o} = 3$ sandwiched between two fully-connected layers. The first two elements of the intermediate representation are each acted on by activation functions $f_{i}$, while the remaining two elements are multiplied together.}
\label{fig:EQLLayer}
\end{center}
\end{figure}

The second stage takes this intermediate representation $z$ to the final $n_{o}$-dimensional output $y$ via a non-linear transformation. For this transformation, the elements of $z$ are divided into two parts which are acted upon differently. To $u$ of the elements, an activation function is applied. These nodes are called unary units. In principle, a different activation function can be applied to each unary unit. The other $v$ elements are pairwise multiplied, giving $\frac{v}{2}$ output values. These nodes are called binary units. This pairwise multiplication step is the key aspect of the layer, as it allows for nonlinear interactions between the nodes of the network. More generally, nonlinear interactions between the nodes other than pairwise multiplication could also be utilized. A diagrammatic representation of an EQL layer is shown in Figure~\ref{fig:EQLLayer}.

\subsection{Model selection}

Our final neural network model is simple, with only one hidden EQL layer between the input and output layers. The input layer consists of 22 nodes as previously described. The linear stage of the EQL layer contains 45 nodes. Of these, 15 are unary units, with the other 30 being the binary units. The output from this layer thus consists of $15 + \frac{30}{2} = 30$ nodes. The unary activation which was found to be the most successful was to square each of the unary units. The output stage consisted of one node, whose value represents the natural logarithm of the number of FRTs. This final output had a ReLU activation function to ensure that the output was nonnegative.

The Adam optimizer was used, with hyperparameters $\beta_{1}=0.9, \beta_{2}=0.99, \textrm{and }\epsilon=1\times10^{-8}$. At each layer, an L1 regularizer with $\lambda=0.001$ was used on all weights in order to select for the most important features. In addition, a dropout rate of 0.1 was used for the EQL hidden layer to help select the most optimal neuron configuration. 

To select a model, we trained across multiple ranges $\h11_{min} \le \h11 \le \h11_{max}$ and examined the results. As the facets at $\h11 \le 5$ have few triangulations and constitute relatively few data points, we chose $\h11_{min} \ge 6$ in each case. In order to have an adequate range on which to test our model's extrapolation, we also chose $\h11_{max} \le 11$.

The training data sets consisted of an equal number of randomly chosen data points from each $\h11$ value. This was done rather than taking a percentage of the data to avoid biasing the model towards fitting to the higher $\h11$ values; there are, for example, over 37 times as many facets at $\h11=11$ as at $\h11=6$. We then tested each model on the full set of values across its training range to see how it performed. Our metric of choice was the mean absolute percentage error (MAPE), as this gives a size-independent measurement of how well the model is performing. The results of our training are shown in Table~\ref{table:ModelSelection}.

\begin{table}[t]
\centering
\begin{tabular}{|c|c|c|c|c|c|c|}
\cline{4-6}
\multicolumn{3}{c}{} & \multicolumn{3}{|c|}{\textbf{Extrapolation MAPE}} \\ \hline
\textbf{$\h11_{min}$} & \textbf{$\h11_{max}$} & \textbf{Test MAPE} & \textbf{$\h11=12$} & \textbf{$\h11=13$} & \textbf{$\h11=14$} \\ \hline
6 & 10 & 7.297 & 6.647 & 6.699 & 6.598 \\ \hline
7 & 10 & 6.001 & 7.512 & 7.626 & 7.469 \\ \hline
8 & 10 & 7.184 & 5.048 & 5.172 & 5.834 \\ \hline
6 & 11 & 5.643 & 4.393 & 4.490 & 4.416 \\ \hline
7 & 11 & 6.967 & 7.512 & 7.626 & 7.469 \\ \hline
8 & 11 & 5.551 & 4.444 & 4.463 & 4.934 \\ \hline
\end{tabular}
\caption{Results of training our model on various $\h11$ ranges. The model with $\h11_{min}=6$,  $\h11_{max}=11$ performs well on the test set and the best on extrapolation to higher $\h11$ values.}
\label{table:ModelSelection}
\end{table}

As the table shows, the model trained on the largest range of $\h11$ values, from 6 to 11, performed the best on both the test set and when extrapolating to higher values. This is the model that we used to generate predictions for the rest of the data set. We note that the extrapolation range contains a large number of points: there are 92162, 108494, and 124700 triangulated facets at $\h11=12, 13, \textrm{ and } 14$, respectively. We also note that, regardless of the training range, the models have stable MAPE values in the extrapolation region. 

\begin{figure}
\begin{center}
\begin{subfigure}{.3\textwidth}
	\includegraphics[scale=0.27]{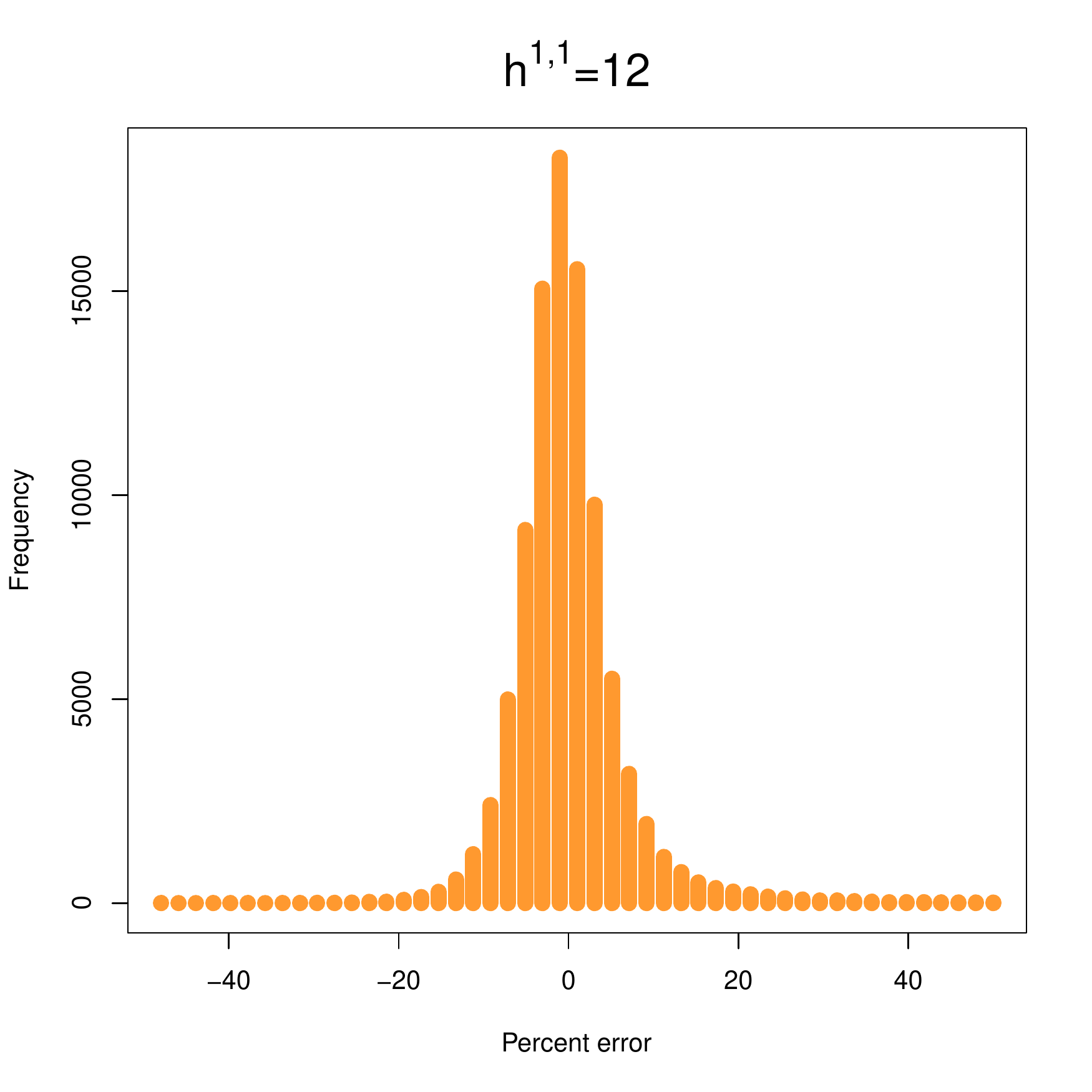}
\end{subfigure}
\begin{subfigure}{.3\textwidth}
	\includegraphics[scale=0.27]{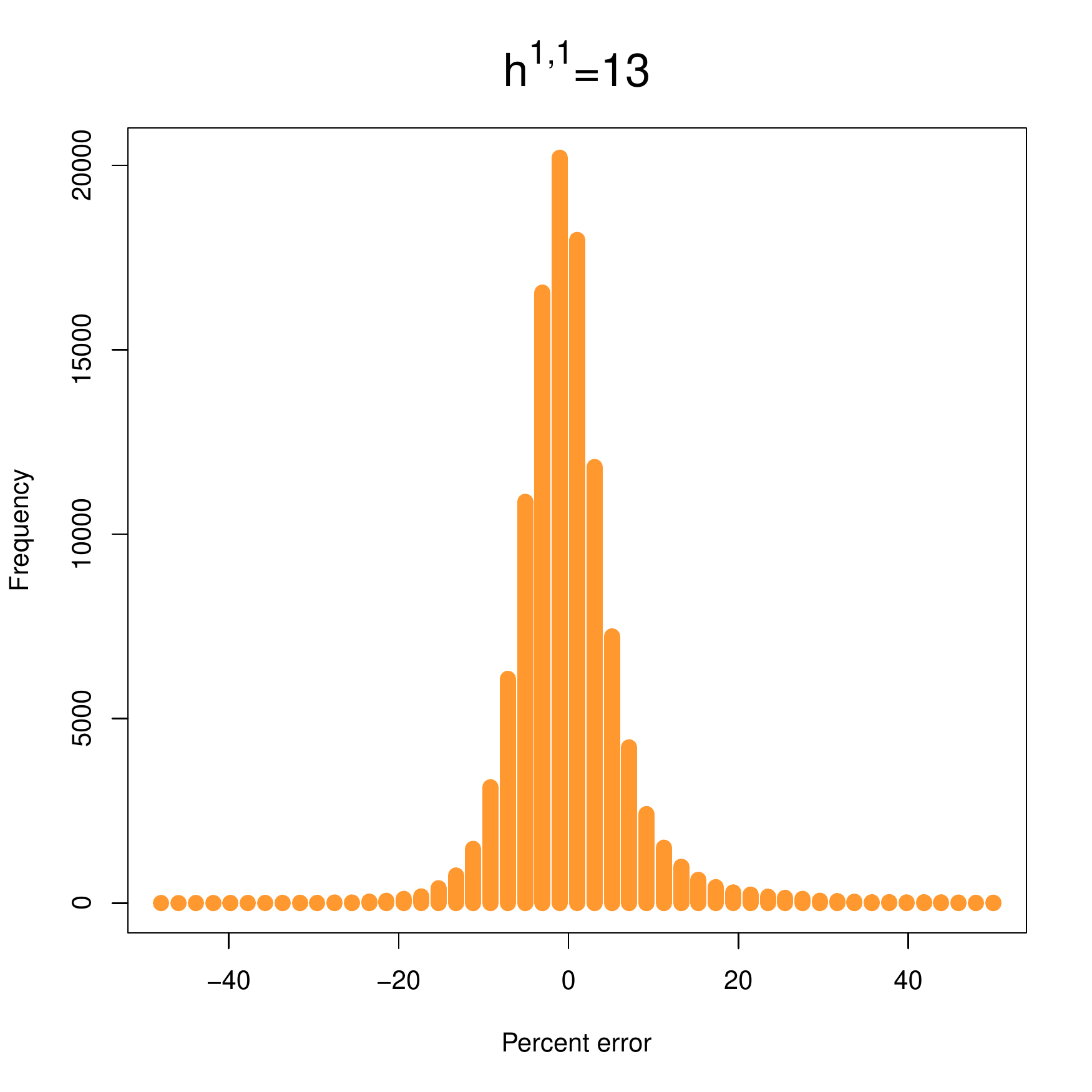}
\end{subfigure}
\begin{subfigure}{.3\textwidth}
	\includegraphics[scale=0.27]{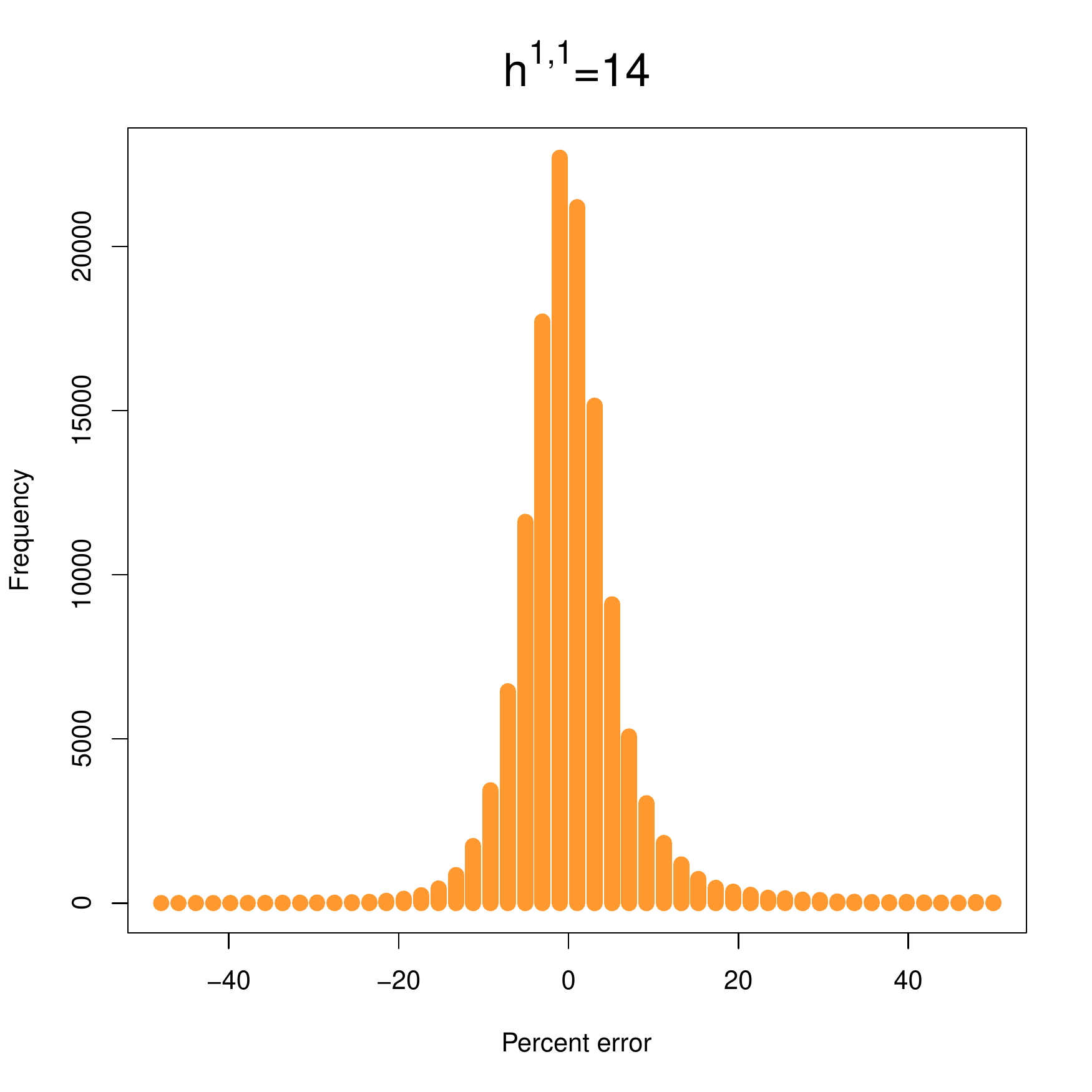}
\end{subfigure}
\end{center}
\caption{Histograms of the percent error of our chosen model's predictions in the extrapolation region. Compare to extrapolations in the EQL network in Figure \ref{fig:PEhistograms}.}
\label{fig:PEhistogramsEQL}
\end{figure}

One might worry that, despite the promising MAPE values, this model is persistently underpredicting like the previous model. However, examining the mean predicted values (Table \ref{table:ModelMeans}) and the distributions of percent errors (Figure \ref{fig:PEhistogramsEQL}), we can see that this is not the case. The mean predicted values stay close to the true means, and the percent errors stay centered around zero as $\h11$ increases.

\begin{table}
\centering
\begin{tabular}{|c|c|c|}
\hline
\textbf{$\h11$} & \textbf{Mean value} & \textbf{Predicted mean} \\ \hline
12 & 10.733 & 10.722 \\ \hline
13 & 11.755 & 11.591 \\ \hline
14 & 12.638 & 12.492 \\ \hline
\end{tabular}
\caption{The true mean values and the mean predicted by our model in the extrapolation region}
\label{table:ModelMeans}
\end{table}

On the other hand, the
extrapolation here is only to three $\h11$ values that are not
in the training set, which provides some cause for concern since we
will be interested in the polytope with maximal $\h11$, which has $\h11=491$.
For this reason we will perform an analogous analysis of $3d$ polytopes
in Section $\ref{sec:3d}$, and will demonstrate that in that case the EQL
accurately predicts the number of FRTs for $\h11$ values significantly beyond
 that in the training set; e.g., training up to $\h11=11$, we will accurately predict numbers of FRTs for all available facets, which go up to $\h11=25$. 
 This suggests that the EQL may also make accurate predictions in the
 4d polytope case at values of $\h11$ significantly beyond the training set.
 
For completeness, we present in Figure \ref{fig:FacetFormula4d} the formula learned by our EQL model. The variables $x_{i}$ are in the same order as in the example in Section 4.3. This formula does not seem to lend itself to any intuitive interpretation. Its interpretability is also hindered by the fact that its variables will generally be of different scales. To attempt to adjust for this, we normalized each variable so that it has an expectation value of 1. This was done using the mean value of each variable in the training data. We then calculated $\textrm{log}_{10}(|c|)$ for each coefficient $c$ in the normalized expression to determine their relative importance on the value. The results of this are shown in Figure \ref{fig:4dHeatmap}. We can see that there are two brightest squares, which correspond to the interactions of the number of unique 1-simplices in the seed triangulation with both the number of 3-simplices, and the number of 1-simplices without accounting for redundancy. However, while these terms have the highest values, there are many others at a similar order of magnitude. Thus, while we can say that some terms are more important than others, the complicated form of the function, along with a lack of knowledge of how these variables will change at larger $\h11$, makes any stronger statements difficult.

\begin{figure}
$\textrm{ln}(N_{FRT}) = -0.425 x_{0}^{2} + 0.29447 x_{0} x_{1} - 0.2304 x_{1}^{2} + 0.02462 x_{0} x_{2} - 0.17529 x_{1} x_{2} - 0.3368 x_{2}^{2} + 0.72012 x_{0} x_{3} + 0.0707 x_{1} x_{3} + 0.00583 x_{2} x_{3} - 0.40825 x_{3}^{2} - 0.01146 x_{0} x_{4} - 0.00008 x_{1} x_{4} - 0.0789 x_{2} x_{4} + 0.00599 x_{3} x_{4} - 0.02246 x_{4}^{2} + 0.35742 x_{0} x_{5} + 0.00696 x_{1} x_{5} + 0.39255 x_{2} x_{5} - 0.35135 x_{3} x_{5} + 0.09 x_{4} x_{5} - 0.2482 x_{5}^{2} - 0.19063 x_{0} x_{6} - 0.02357 x_{1} x_{6} - 0.08904 x_{2} x_{6} + 0.20651 x_{3} x_{6} - 0.04321 x_{4} x_{6} + 0.21098 x_{5} x_{6} - 0.0609 x_{6}^{2} + 0.20861 x_{0} x_{7} + 0.05763 x_{1} x_{7} + 0.5381 x_{2} x_{7} - 0.19461 x_{3} x_{7} + 0.00197 x_{4} x_{7} - 0.41043 x_{5} x_{7} + 0.12497 x_{6} x_{7} - 0.15195 x_{7}^{2} + 0.02359 x_{0} x_{8} + 0.0095 x_{1} x_{8} + 0.05497 x_{2} x_{8} - 0.03016 x_{3} x_{8} + 0.00692 x_{4} x_{8} - 0.10282 x_{5} x_{8} + 0.06078 x_{6} x_{8} + 0.00153 x_{7} x_{8} + 0.00301 x_{8}^{2} - 0.29528 x_{0} x_{9} - 0.00072 x_{1} x_{9} - 0.20046 x_{2} x_{9} + 0.2274 x_{3} x_{9} - 0.01136 x_{4} x_{9} + 0.25023 x_{5} x_{9} - 0.13467 x_{6} x_{9} + 0.23766 x_{7} x_{9} + 0.08246 x_{8} x_{9} - 0.11071 x_{9}^{2} + 0.02683 x_{0} x_{10} - 0.00254 x_{1} x_{10} + 0.02553 x_{2} x_{10} + 0.01278 x_{3} x_{10} + 0.00554 x_{4} x_{10} - 0.04584 x_{5} x_{10} + 0.03714 x_{6} x_{10} + 0.01352 x_{7} x_{10} + 0.00105 x_{8} x_{10} + 0.20665 x_{9} x_{10} - 0.00099 x_{10}^{2} - 0.34428 x_{0} x_{11} + 0.05335 x_{1} x_{11} - 0.00715 x_{2} x_{11} + 0.33024 x_{3} x_{11} - 0.01736 x_{4} x_{11} + 0.16713 x_{5} x_{11} - 0.08848 x_{6} x_{11} + 0.08456 x_{7} x_{11} + 0.03144 x_{8} x_{11} - 0.11092 x_{9} x_{11} - 0.00151 x_{10} x_{11} - 0.07556 x_{11}^{2} + 0.02349 x_{0} x_{12} + 0.0031 x_{1} x_{12} - 0.01155 x_{2} x_{12} - 0.03125 x_{3} x_{12} - 0.00223 x_{4} x_{12} + 0.01197 x_{5} x_{12} - 0.00373 x_{6} x_{12} - 0.0372 x_{7} x_{12} + 0.02864 x_{8} x_{12} - 0.07238 x_{9} x_{12} + 0.01227 x_{10} x_{12} + 0.0125 x_{11} x_{12} - 0.00375 x_{12}^{2} - 0.00624 x_{0} x_{13} + 0.00116 x_{1} x_{13} + 0.02883 x_{2} x_{13} - 0.00045 x_{3} x_{13} - 0.00025 x_{4} x_{13} - 0.02332 x_{5} x_{13} + 0.00198 x_{6} x_{13} - 0.0065 x_{7} x_{13} - 0.07052 x_{8} x_{13} + 0.04967 x_{9} x_{13} - 0.04768 x_{10} x_{13} - 0.00442 x_{11} x_{13} + 0.01206 x_{12} x_{13} + 0.00333 x_{13}^{2} - 0.0373 x_{0} x_{14} - 0.01724 x_{1} x_{14} + 0.00677 x_{2} x_{14} + 0.08978 x_{3} x_{14} + 0.02568 x_{4} x_{14} + 0.03319 x_{5} x_{14} - 0.01481 x_{6} x_{14} + 0.00051 x_{7} x_{14} - 0.00392 x_{8} x_{14} - 0.06852 x_{9} x_{14} - 0.05699 x_{10} x_{14} - 0.0266 x_{11} x_{14} + 0.02735 x_{12} x_{14} - 0.00365 x_{13} x_{14} + 0.00757 x_{14}^{2} - 0.07426 x_{0} x_{15} - 0.07933 x_{1} x_{15} - 0.00701 x_{2} x_{15} + 0.11551 x_{3} x_{15} + 0.01621 x_{4} x_{15} + 0.0459 x_{5} x_{15} - 0.03323 x_{6} x_{15} + 0.01862 x_{7} x_{15} + 0.00938 x_{8} x_{15} - 0.03075 x_{9} x_{15} - 0.00161 x_{10} x_{15} - 0.03737 x_{11} x_{15} + 0.00556 x_{12} x_{15} - 0.00167 x_{13} x_{15} - 0.01574 x_{14} x_{15} - 0.01246 x_{15}^{2} - 0.07638 x_{0} x_{16} + 0.02935 x_{1} x_{16} + 0.08977 x_{2} x_{16} + 0.01429 x_{3} x_{16} + 0.02235 x_{4} x_{16} - 0.06464 x_{5} x_{16} + 0.03652 x_{6} x_{16} + 0.03147 x_{7} x_{16} + 0.00244 x_{8} x_{16} + 0.00924 x_{9} x_{16} - 0.01379 x_{10} x_{16} - 0.00638 x_{11} x_{16} + 0.02373 x_{12} x_{16} - 0.03973 x_{13} x_{16} - 0.00033 x_{14} x_{16} + 0.00165 x_{15} x_{16} - 0.0025 x_{16}^{2} + 0.16209 x_{0} x_{17} + 0.00283 x_{1} x_{17} + 0.00744 x_{2} x_{17} - 0.18455 x_{3} x_{17} + 0.00375 x_{4} x_{17} - 0.08754 x_{5} x_{17} + 0.05244 x_{6} x_{17} - 0.04801 x_{7} x_{17} - 0.02091 x_{8} x_{17} + 0.10988 x_{9} x_{17} - 0.00019 x_{10} x_{17} + 0.05868 x_{11} x_{17} - 0.00919 x_{12} x_{17} + 0.00468 x_{13} x_{17} - 0.00073 x_{14} x_{17} + 0.02233 x_{15} x_{17} - 0.01548 x_{16} x_{17} - 0.02151 x_{17}^{2} + 0.21655 x_{0} x_{18} + 0.00257 x_{1} x_{18} - 0.00186 x_{2} x_{18} - 0.16816 x_{3} x_{18} - 0.0052 x_{4} x_{18} - 0.06696 x_{5} x_{18} + 0.05542 x_{6} x_{18} - 0.07719 x_{7} x_{18} - 0.01278 x_{8} x_{18} + 0.04658 x_{9} x_{18} - 0.11095 x_{10} x_{18} + 0.07151 x_{11} x_{18} + 0.02977 x_{12} x_{18} + 0.0495 x_{14} x_{18} + 0.02117 x_{15} x_{18} + 0.0208 x_{16} x_{18} - 0.07068 x_{17} x_{18} - 0.01711 x_{18}^{2} - 0.04685 x_{0} x_{19} + 0.12461 x_{1} x_{19} - 0.00152 x_{2} x_{19} + 0.00549 x_{3} x_{19} - 0.0037 x_{4} x_{19} + 0.02334 x_{5} x_{19} - 0.00204 x_{6} x_{19} + 0.01064 x_{7} x_{19} - 0.00468 x_{8} x_{19} - 0.01531 x_{9} x_{19} - 0.00025 x_{10} x_{19} - 0.02014 x_{11} x_{19} + 0.00253 x_{12} x_{19} + 0.00039 x_{13} x_{19} + 0.01247 x_{14} x_{19} + 0.00623 x_{15} x_{19} - 0.01978 x_{16} x_{19} + 0.01116 x_{17} x_{19} + 0.00997 x_{18} x_{19} - 0.00125 x_{19}^{2} - 0.24571 x_{0} x_{20} + 0.01527 x_{1} x_{20} + 0.01513 x_{2} x_{20} + 0.21444 x_{3} x_{20} - 0.00042 x_{4} x_{20} + 0.01284 x_{5} x_{20} + 0.01556 x_{6} x_{20} + 0.10175 x_{7} x_{20} + 0.01378 x_{8} x_{20} + 0.01132 x_{9} x_{20} - 0.00014 x_{10} x_{20} - 0.08197 x_{11} x_{20} + 0.05976 x_{12} x_{20} - 0.09782 x_{13} x_{20} - 0.0311 x_{14} x_{20} - 0.02498 x_{15} x_{20} - 0.01261 x_{16} x_{20} + 0.04463 x_{17} x_{20} + 0.03959 x_{18} x_{20} - 0.01115 x_{19} x_{20} - 0.02539 x_{20}^{2} - 0.00024 x_{0} x_{21} + 0.01386 x_{1} x_{21} + 0.02516 x_{2} x_{21} + 0.00222 x_{3} x_{21} + 0.00004 x_{4} x_{21} - 0.06183 x_{5} x_{21} + 0.04606 x_{6} x_{21} + 0.01789 x_{7} x_{21} - 0.01325 x_{8} x_{21} + 0.02409 x_{9} x_{21} + 0.00023 x_{10} x_{21} + 0.00006 x_{11} x_{21} - 0.00317 x_{12} x_{21} + 0.00541 x_{13} x_{21} - 0.00012 x_{14} x_{21} + 0.00012 x_{15} x_{21} - 0.00592 x_{16} x_{21} + 0.02185 x_{17} x_{21} - 0.00001 x_{18} x_{21} - 0.00003 x_{20} x_{21} + 0.0257 x_{21}^{2} - 1.548990 x_{0} + 3.819380 x_{1} + 4.156280 x_{2} - 1.006350 x_{3} + 1.110140 x_{4} - 2.591030 x_{5} + 1.540610 x_{6} - 2.829660 x_{7} - 1.5943 x_{8} + 1.383880 x_{9} - 2.551 x_{10} + 0.18922 x_{11} + 0.72398 x_{12} + 0.04856 x_{13} + 0.48053 x_{14} + 0.6916 x_{15} - 1.253460 x_{16} - 0.70274 x_{17} + 0.44341 x_{18} - 1.167680 x_{19} - 1.8572 x_{20} - 1.170990 x_{21} - 23.70712$
\caption{The formula learned by our model for the number of FRTs of a facet. The meanings of the $x_{i}$ are described in Section 4.3.}
\label{fig:FacetFormula4d}
\end{figure}

\begin{figure}
\begin{center}
\includegraphics[scale=0.6]{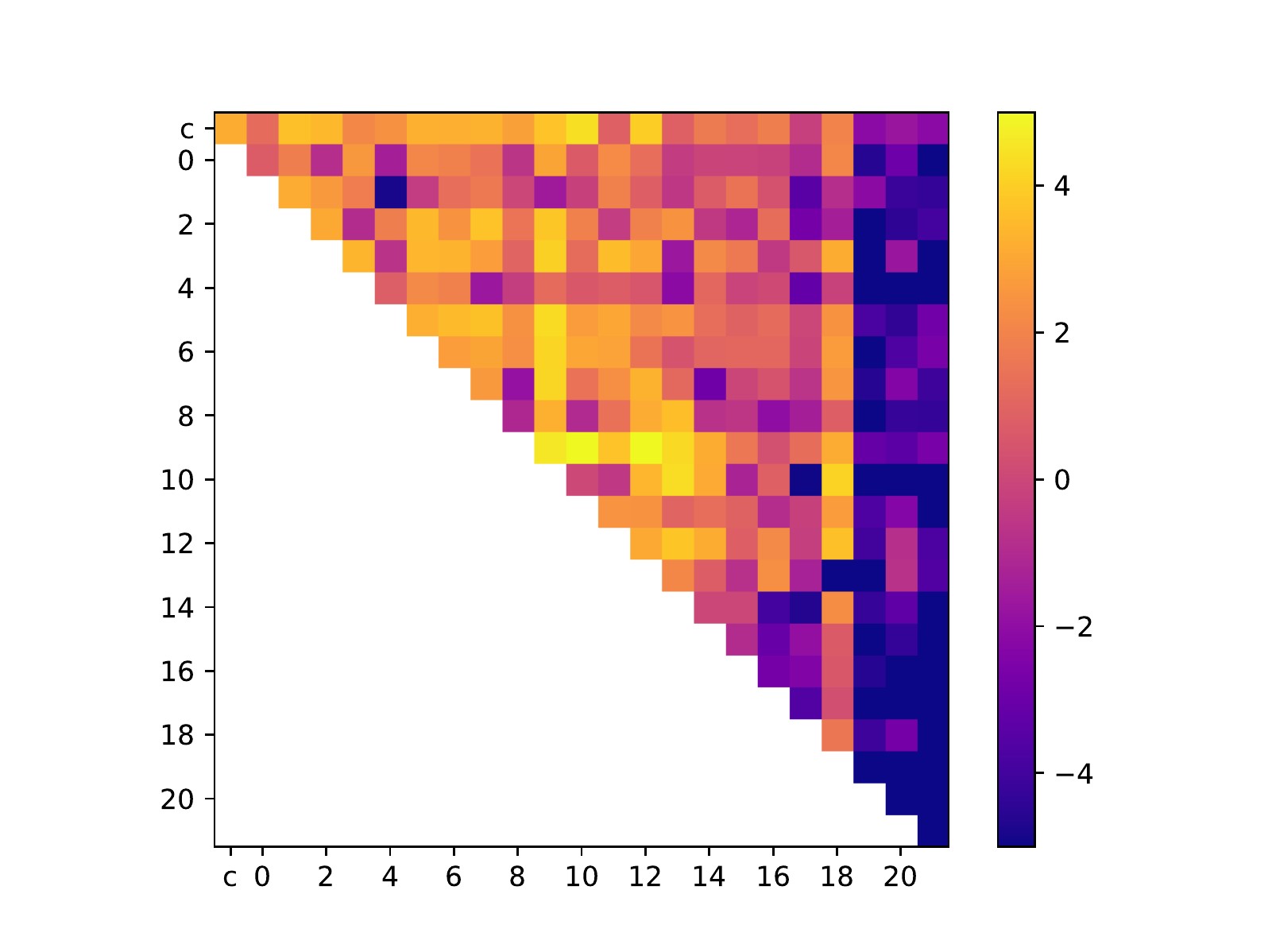}
\caption{A heatmap showing $\textrm{log}_{10}(|c|)$ for each coefficient $c$ in the formula, after rescaling each variable to have expectation value 1. The top row corresponds to the constant term (top left square) and the linear terms.}
\label{fig:4dHeatmap}
\end{center}
\end{figure}

\subsection{Estimate for the total upper bound}

Having obtained a satisfactory model, we generated the necessary input data for all of the facets. We then fed the data into the neural network to obtain the estimated number of FRTs for each facet.

Naively, for a given reflexive polytope, its FRSTs would be given by the set of all possible combinations of its facets' FRTs. But in reality, the number of FRSTs is reduced by two considerations. The first is that given two 3d facets $F_{1}$ and $F_{2}$, the triangulation of each induces a triangulation on the intersection $F_{1}\bigcap F_{2}$, and these induced triangulations may not overlap. The second consideration is that even if the induced triangulations do overlap, the aggregated triangulation of the reflexive polytope may fail to be regular even though the individual facet FRTs are regular.
Thus, using the estimated number of FRTs for each facet, we are only able to estimate an upper bound for each 4d reflexive polytope. So, for a given reflexive polytope $\Delta$ with facets $F_{i}$, we know that
\begin{equation}
N_{FRST}(\Delta) \le \prod_{i}N_{FRT}(F_{i})\, ,
\label{NFRST}
\end{equation}
where $N_{FRST}$ and $N_{FRT}$ are the number of FRSTs and FRTs of the 2-skeleton of a given reflexive polytope and facet, respectively. Via this inequality, the upper bound is given by the product of the facet FRTs. One can then calculate the estimate for each 4d polytope and sum the results. Determining the degree to which the product of FRTs overestimates the number of FRSTs is left to a future work.

In practice, since our neural network predicted the natural logarithm of the number of FRSTs, we calculated the estimate for each polytope as
\begin{equation}
 N_{FRST}(\Delta) = \textrm{exp}\big(\sum_{i} \textrm{ln}(N_{FRT}(F_{i})\big)\, .
 \label{NFRSTactual}
\end{equation}
In these calculations, we used the actual value for $\ln(N_{FRT}(F_{i}))$ whenever it was known, and the neural network prediction otherwise. As noted in Section~\ref{sec:class}, the triangulated facets account for over 88\% of all facets.

Performing this calculation for each polytope, it was found that the single polytope whose dual has $\h11=491$ (the greatest $\h11$ value) dominated the count. Our estimation method predicts that it has $1.465 \times 10^{10,505}$ FRSTs. This is more than any other polytope by over 1300 orders of magnitude, and so this value effectively serves as the total triangulation number for the entire set. Considering that this polytope has 680 integral points, 40 more than any other, it is unsurprising that our method has identified this polytope as possessing the most triangulations.

\subsection{The $\h11=491$ polytope}

The polytope whose FRST count dominates the database is the polytope dual to the single $\h11=491$ polytope, which we will call $\D491$. This polytope has 680 integral points and five facets, of which only four are unique. The polytope and its four facets $F_{i}$ are given by the convex hulls
\begin{eqnarray}
\D491 &=& \textrm{conv}(\{\{1,0,0,0\},\{0,1,0,0\},\{0,0,1,0\},\{21,28,36,42\},\{-63,-56,-48,-42\}\})
\nonumber \\
F_{1} &=& \textrm{conv}(\{\{1,0,0,0\},\{0,1,0,0\},\{0,0,1,0\},\{21,28,36,42\}\}) \nonumber \\
F_{2} &=& \textrm{conv}(\{\{1,0,0,0\},\{0,1,0,0\},\{3,4,6,0\},\{3,4,6,84\}\}) \\
F_{3} &=& \textrm{conv}(\{\{1,0,0,0\},\{0,1,0,0\},\{7,8,14,0\},\{7,8,14,84\}\}) \nonumber \\
F_{4} &=& \textrm{conv}(\{\{1,0,0,0\},\{0,1,0,0\},\{7,15,21,0\},\{7,15,21,84\}\})\, . \nonumber
\end{eqnarray}
The facet $F_{1}$ first appears as a dual facet at $\h11=23$, and appears twice in $\D491$. The facets $F_{2}, F_{3}$ and $F_{4}$ each appear once in $\D491$ and nowhere else in the database.

Our model predicts that $F_{1}$ will have $e^{29.32} = 5.41 \times 10^{12}$ FRTs. As it appears twice, it will contribute $e^{2\times29.32} = 2.93 \times 10^{25}$ to the triangulation number for this polytope, which is a subdominant contribution. $F_{2}$ has a larger contribution, as our model predicts that it will have $e^{2391.5} = 4 \times 10^{1038}$ FRTs. However, the triangulation number for the polytope is dominated by $F_{3}$ and $F_{4}$, which our model predicts to have $e^{10,753.0} = 1 \times 10^{4670}$ and $e^{10,985.9} = 1.25 \times 10^{4771}$ FRTs, respectively. Combining these, we get the estimate of the number of FRSTs for this polytope:
\begin{equation} 
N_{FRST}(\D491) = (2.93 \times 10^{25})(4 \times 10^{1038})(1 \times 10^{4670})(1.25 \times 10^{4771}) = 1.5 \times 10^{10,505}\, .
\label{N491}
\end{equation}
Finally, we note that according to our model, the facets $F_{4}$ and $F_{3}$ have the most and second-most FRTs among all facets. 

To estimate a range of uncertainty for this value, we can use the extrapolated MAPE values from our model in Table \ref{table:ModelSelection}. For $\h11=12,13,14$, our model had MAPE values of $4.393, 4.490$, and $4.416$. Using the average of these values, $4.433$, as the percent error for each $\textrm{ln}(N_{FRT})$ estimate in $\D491$, we have
\[ F_{1} : \textrm{ln}(N_{FRT}) = 29.32 \pm 1.30,\]
\[ F_{2} : \textrm{ln}(N_{FRT}) = 2391.5 \pm 106.0,\]
\[ F_{3} : \textrm{ln}(N_{FRT}) = 10753.0 \pm 476.7,\]
\[ F_{4} : \textrm{ln}(N_{FRT}) = 10985.9 \pm 487.0.\]
Propagating these errors, we find that for $\D491$ we have $\textrm{log}_{10}(N_{FRT}) = 10,505.2 \pm 292.6$. This gives a range for our estimate of $10^{10,505.2\pm292.6}=[10^{10,212.6}, 10^{10,797.8}]$.

\section{Comparison to 3d Polytopes\label{sec:3d}}

The biggest point of uncertainty involving our analysis of the 4d reflexive polytopes is the inability to test our model on facets that first appear beyond $\h11=15$. As all evidence suggests that these facets - and in particular those that first appear at high $\h11$ values - will be responsible for the dominant contribution to the total number of FRSTs, it would be preferable to have some confidence in our model's accuracy in this region.

Though obtaining the true number of FRTs for these largest facets is currently infeasible, we hope to gain confidence in our model by showing that we can achieve the same goal for the 2d facets of the 3d reflexive polytopes. As these facets are smaller, the number of FRTs for the majority of the facets can be triangulated. Therefore a model's extrapolation ability can be more thoroughly tested. Specifically,
we will demonstrate that EQLs trained to
predict $\textrm{ln}(N_{FRT})$ predictions
on data with relatively low $h^{1,1}$ nevertheless accurately extrapolate to much higher $h^{1,1}$ values, lending some credence
to our above assumption the 4d case.

\subsection{Classification of the 2d facets}

As in Section 2, we began by classifying the 2d facets of the 3d reflexive polytopes to avoid redundancy in our calculations. The same classification method as in Section 2 - for each facet, we obtained a 3d polytope by attaching the origin, calculated the normal form, and removed the origin afterwards.

In Section 2 we organized the facets by the first $\h11$ value at which they appeared. As $\h11$, as obtained by Batyrev's formula, is equal to $20$ for every 3d reflexive polytope, we instead organized the facets by the number of integral points $|\Delta|$ in the smallest polytope $\Delta$ in which they appeared. We note that this quantity is equal to $\h11(B)+4$, where $B$ is the smooth weak Fano toric threefold associated to $\Delta$. For the remainder of this section, by $\h11$ we will mean $\h11(B) = |\Delta| - 4$.

As in Section 2, the calculations were performed using our \cpp lattice polytope implementation. Due to the small size of the 3d reflexive dataset, the classification was completed in just over 20 seconds.

\begin{center}
\begin{figure}[th!]
\begin{subfigure}{.5\textwidth}
	\includegraphics[scale=0.45]{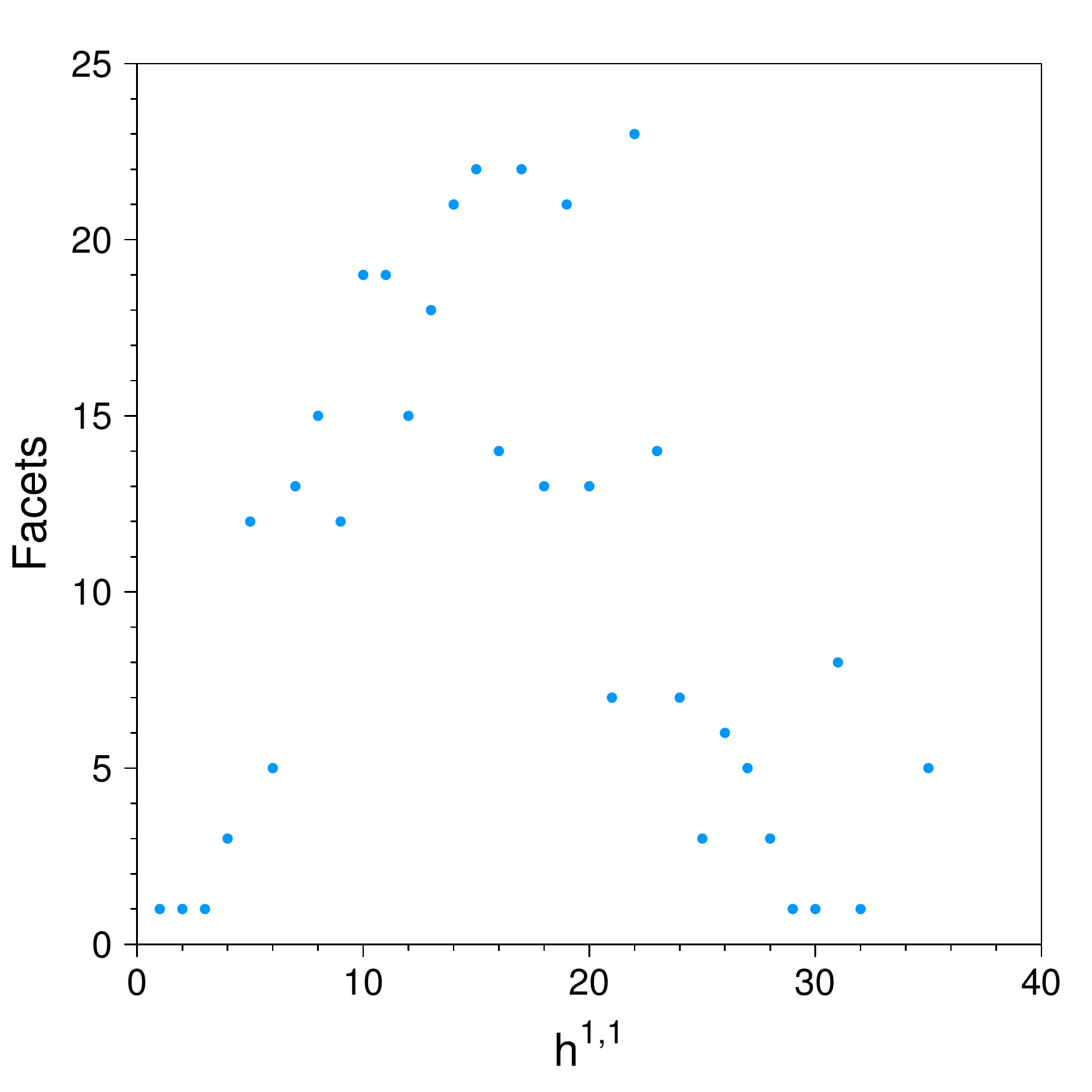}
\end{subfigure}
\begin{subfigure}{.5\textwidth}
	\includegraphics[scale=0.45]{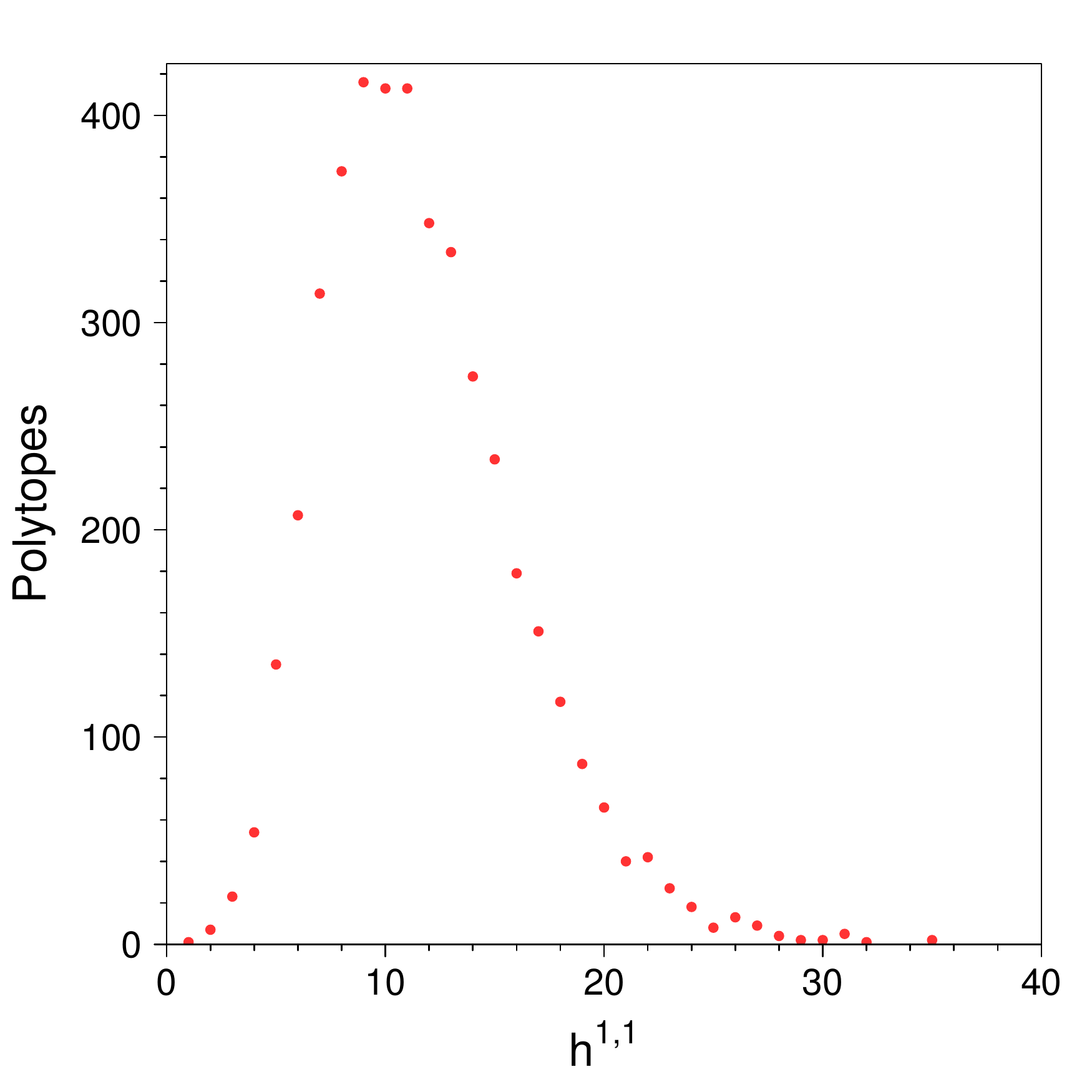}
\end{subfigure}
\caption{\textbf{(Left)} The number of new facets at each $\h11$ value. \textbf{(Right)} The number of reflexive polytopes at each $\h11$ value.}
\label{fig:3DDistributions}
\end{figure}
\end{center}

Performing our classification, we found that the $4,319$ 3d reflexive polytope contain $344$ unique 2d facets. The numbers of facets and polytopes that first appear at each $\h11$ value are shown in Figure \ref{fig:3DDistributions}.

\subsection{Training data}

Having classified the $344$ unique facets, we set out to triangulate as many as possible in order to test our model. We again used {\tt TOPCOM} to obtain the FRT count for each facet. We were able to successfully obtain the number of FRTs for $322$ of the $344$ facets, with all facets that appear up to $\h11=25$ being completed.

Additionally, for 6 of the remaining 12 facets, as well as all of the other $322$, we were able to obtain the number of fine triangulations (FTs). For the $322$ facets for which the number of FRTs was found, there is very close agreement between the numbers of FRTs and FTs. More specifically, the number of FTs was at most 1.026 times the number of FRTs. Thus, we used the number of FTs as an approximation for the number of FRTs for the six additional facets. This gave us values for all facets that appear up through $\h11=30$.

\subsection{Machine learning and extrapolation}

Having triangulated as many facets as possible, we trained models with an EQL hidden layer. Our input data was simpler than for the 3d facets, and consisted of

\begin{itemize}
\item{The number of integral points}
\item{The number of boundary points}
\item{The number of interior points}
\item{The number of vertices}
\item{The length of the longest side}
\item{The length of the shortest side}
\item{The average side length}
\end{itemize}

\begin{figure}[h]
\begin{center}
\includegraphics[scale=0.5]{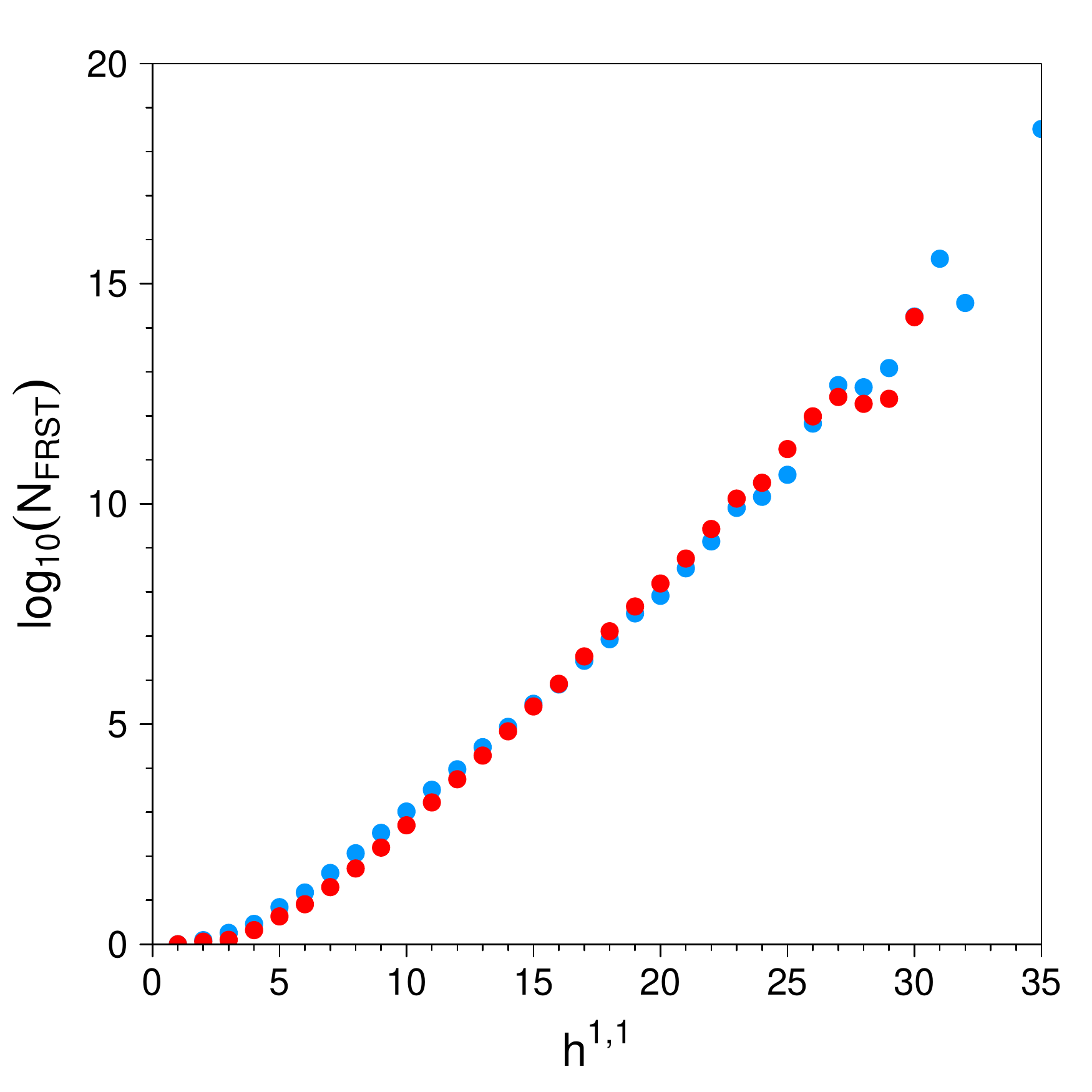}
\end{center}
\caption{The mean value of $\textrm{log}_{10}(N_{FRST})$ at $\h11$, using predicted facet FRT values \textbf{(blue)} and known facet FRT and FT values \textbf{(red)}.}
\label{fig:PolyMeans}
\end{figure}

As before, we had the neural network train on $\textrm{ln}(N_{FRT})$. As the goal of this exercise was to mimic the 4d case, where we only have data for the smallest facets, we restricted our training data to facets that first appear at $4 \le \h11 \le 11$. There are $243$ facets that first appear at $\h11 \ge 12$, meaning that our extrapolation region contains $70.6\%$ of the facets.

Our trained model contains a single EQL layer, which used a combination of linear and quadratic unary activation functions. The performance of the model is shown in Table \ref{table:results}. Recall that $\h11=25$ is the highest value for which all of the FRT values are known; for $26 \le \h11 \le 30$ we approximate the number of FRT by the number of FTs for some facets.

\begin{table}[h]
\begin{center}
\begin{tabular} {|c|c|c|c|}
\hline
\textbf{$\h11$} & \textbf{Facets} & \textbf{MAPE} & \textbf{MSE}\\ \hline
6 & 5 & 7.865 & 0.032 \\
7 & 13 & 16.583 & 0.061 \\
8 & 15 & 8.805 & 0.055 \\
9 & 12 & 5.851 & 0.075 \\
10 & 19 & 5.808 & 0.087 \\
11 & 19 & 10.678 & 0.213 \\
12 & 15 & 8.754 & 0.334 \\
13 & 18 & 9.128 & 0.330 \\
14 & 21 & 10.200 & 0.722 \\
15 & 22 & 8.756 & 0.538 \\
16 & 14 & 9.103 & 0.755 \\
17 & 22 & 9.071 & 0.610 \\
18 & 13 & 7.850 & 0.619 \\
19 & 21 & 10.491 & 1.314 \\
20 & 13 & 10.962 & 2.050 \\
21 & 7 & 9.259 & 1.158 \\
22 & 23 & 9.167 & 0.798 \\
23 & 14 & 14.333 & 6.504 \\
24 & 7 & 7.894 & 1.075 \\
25 & 3 & 2.649 & 0.373 \\
26 & 6 & 12.112 & 2.228 \\
27 & 5 & 3.985 & 0.644 \\
28 & 3 & 6.900 & 0.380 \\
29 & 1 & 5.258 & 0.295 \\
30 & 1 & 2.074 & 0.146 \\
\hline
\end{tabular}
\end{center}
\caption{Performance of our model across $\h11$ of the 3d reflexive polytopes.}
\label{table:results}
\end{table}

As we can see, the model performs well through the extrapolation region of $17 \le \h11 \le 25$. With our trained model, we predicted the number of FRTs for all $344$ facets, and then estimated the number of FRSTs for any given polytope as the product of the predictions for each facet. We also computed these estimates using the real FRT numbers for polytopes where all facets have been triangulated. Taking the mean value at each value of $\h11$, we obtain the graph shown in Figure \ref{fig:PolyMeans}. From this graph we see that the estimates from the model predictions are in good agreement with the estimate made from known values well outside of the training region of $4 \le \h11 \le 11$. In particular, we note that at the highest $\h11$ value available to us, $\h11=30$, our model is in nearly perfect agreement with the true result.

We give below in Figure \ref{fig:FacetFormula3d} the formula learned by our model, with the $x_{i}$ in the order of the list at the beginning of this section. As in the 4d case, we also rescaled the variables to have expectation value 1, with the results being show in Figure \ref{fig3dHeatmap}. Again, we see many warm spots, making further interpretation of the formula difficult.

\begin{figure}
$\textrm{ln}(N_{FRT}) = 0.01418 x_{0}^{2} - 0.03435 x_{0} x_{1} - 0.02165 x_{1}^{2} + 0.11134 x_{0} x_{2} + 0.00201 x_{1} x_{2} + 0.00206 x_{2}^{2} - 0.03566 x_{0} x_{3} - 0.02813 x_{1} x_{3} + 0.00993 x_{2} x_{3} + 0.05023 x_{3}^{2} - 0.03399 x_{0} x_{4} - 0.00929 x_{1} x_{4} - 0.01405 x_{2} x_{4} + 0.11072 x_{3} x_{4} + 0.0694 x_{4}^{2} + 0.04551 x_{0} x_{5} - 0.04939 x_{1} x_{5} + 0.04087 x_{2} x_{5} - 0.00532 x_{3} x_{5} + 0.00719 x_{4} x_{5} - 0.00774 x_{5}^{2} - 0.07105 x_{0} x_{6} + 0.04438 x_{1} x_{6} - 0.11917 x_{2} x_{6} - 0.07082 x_{3} x_{6} - 0.14734 x_{4} x_{6} - 0.007 x_{5} x_{6} + 0.14731 x_{6}^{2} - 0.28707 x_{0} + 0.46716 x_{1} - 0.59766 x_{2} - 0.4975 x_{3} - 0.35609 x_{4} - 0.49381 x_{5} + 1.354040 x_{6} + 5.530190$
\caption{The formula learned by our model for the number of FRTs of a facet. The meanings of the $x_{i}$ are described earlier in this section}
\label{fig:FacetFormula3d}
\end{figure}

\begin{figure}[th]
\begin{center}
\includegraphics[scale=0.6]{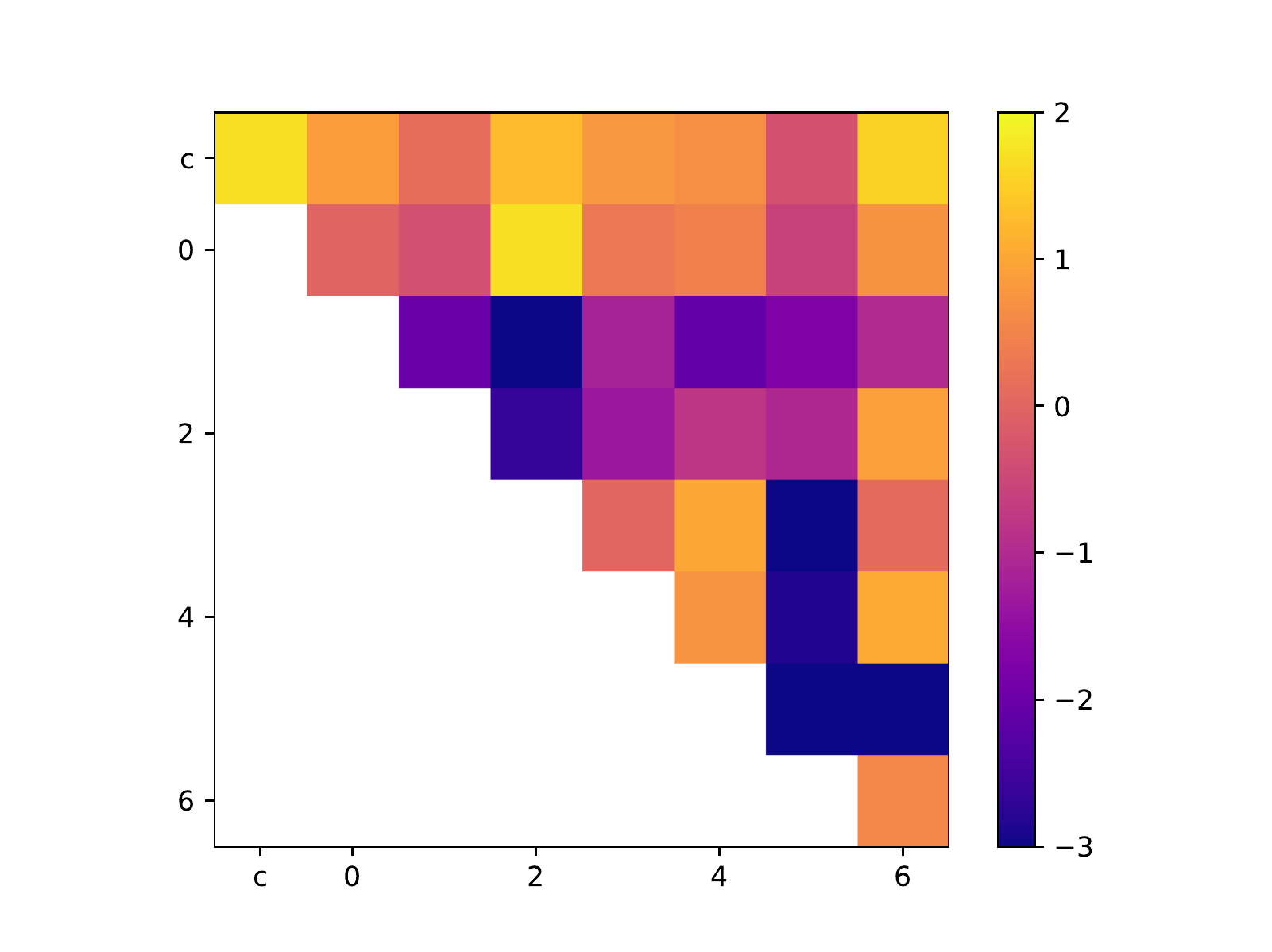}
\caption{A heatmap showing $\textrm{log}_{10}(|c|)$ for each coefficient $c$ in the formula, after rescaling each variable to have expectation value 1. The top row corresponds to the constant term (top left square) and the linear terms.}
\label{fig3dHeatmap}
\end{center}
\end{figure}

%\section{Overcounting}

\section{Conclusion}
\label{sec:conclusion}

In this paper we have provided the first concrete estimate
of the number of fine, regular, star triangulations of 
4d reflexive polytopes,
\begin{equation}
n_{\text{FRST}} \simeq 10^{10,505.2\pm292.6}.
\end{equation}
This provides an upper bound on the number of topologically
distinct Calabi-Yau threefold hypersurfaces in toric varieties.

We attempted a variety of supervised learning techniques for
predicting $n_\text{FRST}$ and found that a neural network
with the equation learning (EQL) architecture performed best.
The estimation was computed by taking products of numbers
of FRTs of facets, where the EQL was trained on 4d polytopes
up to $\h11=11$ and made accurate predictions up to $\h11=14$,
which was the highest $\h11$ at which we were able to compute
FRTs of facets for the sake of validating the trained EQL.
While encouraging that the EQL extrapolated to higher $\h11$,
it is so much smaller than the maximum $\h11=491$ that 
it is hard to trust such a high extrapolation without further
evidence. For that reason, we performed an analogous 
analysis in the case of $3d$ polytopes and found that 
the EQL was able to extrapolate accurately to $\h11=30$,
where the maximum possible value is $\h11=35$, despite
the fact that it was only trained up through $\h11=11$.
This provides some evidence that the corresponding extrapolation in the 4d case
may be trustworthy, despite needing to extrapolate 
far beyond the training region. Indeed, such extrapolations
were part of the motivation for the EQL architecture \cite{1610.02995} in the first place.

After training successful models, we extracted the equations that they learned. The variables were rescaled to have an expectation value of 1 in an attempt to interpret these equations. We found that some variables and terms were more important than others, but the majority of variables made significant contributions, making interpretation of the equation difficult. This may be due to the functional form utilized in our EQLs.

In the process of making the prediction,
we have demonstrated the overall utility of deep neural networks 
in this context, in particular their ability to extrapolate to regions of higher topological complexity. This provides motivation for further studies of triangulations and Calabi-Yau manifolds using techniques from data science.

Though this result is a modest prediction of a single number,
it is a necessary step in understanding the ensemble of
Calabi-Yau threefold hypersurfaces as a whole. We wish to turn 
to refined structures in the ensemble in the future, and
in particular their implications for cosmology. For instance,
studying axion-like particles (ALPs) in this context is particularly
well-motivated since the ensemble is strongly dominated by the polytope that gives
rise to the large number of ALPs.

\section*{Acknowledgements}

We thank Cody Long and Jiahua Tian for useful discussions. B.D.N and J.C are supported by NSF Grant PHY-1620575. J.H. is supported by NSF grant
PHY-1620526.

\pagebreak

\bibliographystyle{unsrt}
\bibliography{refs.bib}
\end{document}